\documentclass[pre,preprint, 10pt, showpacs]{revtex4}
\usepackage{amsmath}

\begin{document}

\title{Chapman-Enskog expansion about nonequilibrium states: the sheared
granular fluid.}
\author{James F. Lutsko}
\affiliation{Center for Nonlinear Phenomena and Complex Systems, Universit\'{e} Libre de
Bruxelles, C.P. 231, Blvd. du Triomphe, 1050 Brussels, Belgium}
\date{\today }

\begin{abstract}
The Chapman-Enskog method of solution of kinetic equations, such as the
Boltzmann equation, is based on an expansion in gradients of the deviations
of the hydrodynamic fields from a uniform reference state (e.g., local
equilibrium). This paper presents an extension of the method so as to allow
for expansions about \emph{arbitrary}, far-from equilibrium reference
states. The primary result is a set of hydrodynamic equations for studying
variations from the arbitrary reference state which, unlike the usual
Navier-Stokes hydrodynamics, does not restrict the reference state in any
way. The method is illustrated by application to a sheared granular gas
which cannot be studied using the usual Navier-Stokes hydrodynamics.
\end{abstract}

\maketitle

\section{Introduction}

The determination of the one-body distribution function, which gives the
probability of finding a particle at some given position,with a given
velocity at a given time, is one of the central problems in nonequilibrium
statistical mechanics. Its time-evolution is in many cases well described by
approximate kinetic equations such as the Boltzmann equation\cite{McLennan},
for low-density gases and the revised Enskog equation\cite{RET},\cite{KDEP},
for denser hard-sphere gases and solids. Only rarely are exact solutions of
these equations possible. Probably the most important technique for
generating approximate solutions to one-body kinetic equations is the
Chapman-Enskog method which, as originally formulated, consists of a
gradient expansion about a local-equilibrium state\cite{ChapmanCowling},\cite%
{McLennan}. The goal in this approach is to construct a particular type of
solution, called a ''normal solution'', in which all space and time
dependence of the one-body distribution occurs implicitly via its dependence
on the macroscopic hydrodynamic fields. The latter are, for a simple fluid,
the density, velocity and temperature fields corresponding to the conserved
variables of particle number, momentum and energy respectively. (In a
multicomponent system, the partial densities are also included.) The
Chapman-Enskog method proceeds to develop the solution perturbatively in the
gradients of the hydrodynamic fields: the distribution is developed as a
functional of the fields and their gradients and at the same time the
equations of motion of the fields, the hydrodynamic equations, are also
developed. The zeroth-order distribution is the local-equilibrium
distribution; at first order, this is corrected by terms involving linear
gradients of the hydrodynamic fields which in turn are governed by the Euler
equations (with an explicit prescription for the calculation of the pressure
from the kinetic theory). At second order, the hydrodynamic fields are
governed by the Navier-Stokes equations, at third order, by the Burnett
equations, etc. The calculations involved in extending the solution to each
successive higher order are increasingly difficult and since the
Navier-Stokes equations are usually considered an adequate description of
fluid dynamics, results above third order (Burnett order) for the Boltzmann
equation and above second (Navier-Stokes) order for the Enskog equation are
sparse. The extension of the Chapman-Enskog method beyond the Navier-Stokes
level is, however, not physically irrelevant since only by doing so is it
possible to understand non-Newtonian viscoelastic effects such as shear
thinning and normal stresses which occur even in simple fluids under extreme
conditions\cite{Lutsko_EnskogPRL},\cite{LutskoEnskog}. 

Recently, interest in non-Newtonian effects has increased because of their
importance in fluidized granular materials. Granular systems are composed of
particles - grains - which lose energy when they collide. As such, there is
no equilibrium state - an undriven homogeneous collection of grains will
cool continuously. This has many interesting consequences such as the
spontaneous formation of clusters in the homogeneous gas and various
segregation phenomena in mixtures\cite{GranularPhysicsToday},\cite%
{GranularRMP},\cite{GranularGases},\cite{GranularGasDynamics}. The
collisional cooling also gives rise to a unique class of nonequilibrium
steady states due to the fact that the cooling can be balanced by the
viscous heating that occurs in inhomogeneous flows. One of the most widely
studied examples of such a system is a granular fluid undergoing planar
Couette flow where the velocity field takes the form $\mathbf{v}\left( 
\mathbf{r}\right) =ay\widehat{\mathbf{x}}$, where $a$ is the shear rate. The
common presence of non-Newtonian effects, such as normal stresses, in these
systems has long been recognized as signalling the need to go beyond the
Navier-Stokes description\cite{SelGoldhirsch}. As emphasized by Santos et al,%
\cite{SantosInherentRheology}, the balance between the velocity gradients,
which determine the rate of viscous heating, and the cooling, arising from a
material property, means that such fluids are inherently non-Newtonian in
the sense that the sheared state cannot be viewed as a perturbation of the
unsheared, homogeneous fluid and so the usual Navier-Stokes equations cannot
be used to study either the rheology or the stability of the sheared
granular fluid. One of the goals of the present work is to show that a more
general hydrodynamic description can be derived for this, and other flow
states, which is able to accurately describe such far-from-equilibrium
states. The formalism developed here is general and not restricted to
granular fluids although they do provide the most obvious application.
Indeed,an application of this form of hydrodynamics has recently been
presented by Garz{\'{o}}\cite{garzo-2005-} who studied the stability of a
granular fluid under strong shear.

The extension of the Chapman-Enskog method to derive the hydrodynamics for
fluctuations about an arbitrary nonequilibrium state might at first appear
trivial but in fact it involves a careful application of the ideas
underlying the method. To illustrate, let $f\left( \mathbf{r},\mathbf{v}%
,t\right) $ be the probability to find a particle at position $\mathbf{r}$
with velocity $\mathbf{v}$ at time $t$. For a $D-$dimensional system in
equilibrium, this is just the (space and time-independent) Gaussian
distribution%
\begin{equation}
f\left( \mathbf{r},\mathbf{v},t\right) =\phi _{0}\left( \mathbf{v}%
;n,T,U\right) =n\left( \frac{m}{2\pi k_{B}T}\right) ^{D/2}\exp \left(
-\left( \mathbf{v}-\mathbf{U}\right) ^{2}/k_{B}T\right)   \label{le}
\end{equation}%
where $n$ is the number density, $k_{B}$ is Boltzmann's constant, $T$ is the
temperature, $m$ is the mass of the particles and $\mathbf{U}$ is the
center-of-mass velocity. The zeroth-order approximation in the
Chapman-Enskog method is the localized distribution $f^{(0)}\left( \mathbf{r}%
,\mathbf{v},t\right) =\phi _{0}\left( \mathbf{v};n\left( \mathbf{r},t\right)
,T\left( \mathbf{r},t\right) ,\mathbf{U}\left( \mathbf{r},t\right) \right) $
or, in other words, the local equilibrium distribution. In contrast, a
homogeneous non-equilibrium steady state might be characterized by some
time-independent distribution 
\begin{equation}
f\left( \mathbf{r},\mathbf{v},t\right) =\Phi _{ss}\left( \mathbf{v};n,T,%
\mathbf{U}\right) 
\end{equation}%
but the zeroth-order approximation in the Chapman-Enskog method will \emph{%
not} in general be the localized steady-state distribution, $f^{(0)}\left( 
\mathbf{r},\mathbf{v},t\right) \neq \Phi _{ss}\left( \mathbf{v};n\left( 
\mathbf{r},t\right) ,T\left( \mathbf{r},t\right) ,\mathbf{U}\left( \mathbf{r}%
,t\right) \right) $. The reason is that a steady state is the result of a
balance - in the example given above, it is a balance between viscous
heating and collisional cooling. Thus, any change in density must be
compensated by, say, a change in temperature or the system is no longer in a
steady state. This therefore gives a relation between density and
temperature in the steady state, say $n=n_{ss}(T)$, so that one has $\Phi
_{ss}\left( \mathbf{v};n,T,\mathbf{U}\right) =\Phi _{ss}\left( \mathbf{v}%
;n_{ss}(T),T,\mathbf{U}\right) $. Clearly, it makes no sense to simply
''localize'' the hydrodynamic variables as the starting point of the
Chapman-Enskog method since, in a steady state, the hydrodynamic variables
are not independent. Limited attempts have been made in the past to perform
the type of generalization suggested here. In particular, Lee and Dufty
considered this problem for the specific case of an ordinary fluid under
shear with an artificial thermostat present so as to make possible a steady
state\cite{MirimThesis},\cite{MirimThesisArticle}. However, the issues
discussed in this paper were circumvented through the use of a very
particular type of thermostat so that, while of theoretical interest, that
calculation cannot serve as a template for the more general problem.

In Section II, the abstract formulation of the Chapman-Enskog expansion for
fluctuations about a non-equilibrium state is proposed. It not only requires
care in understanding the zeroth order approximation, but a generalization
in the concept of a normal solution. In Section III, the method is
illustrated by application to a simple kinetic theory for a sheared granular
gas. Explicit expressions are given for the full complement of transport
coefficients. One unique feature of the hydrodynamics obtained in this case
is that several transport coefficients depend linearly on fluctuations in
the velocity in the $y$-direction (i.e. in the direction of the velocity
gradient). The section concludes with a brief summary of the resulting
hydrodynamics and of the linearized form of the hydrodynamic equations which
leads to considerable simplification. The paper ends in Section IV with a
summary of the results, a comparison of the present results to the results
of the standard Chapman-Enskog analysis and a discussion of further
applications.

\section{The Chapman-Enskog expansion about an arbitrary state}

\subsection{Kinetic theory}

Consider a single-component fluid composed of particles of mass $m$ in $D$
dimensions. In general, the one-body distribution will obey a kinetic
equation of the form%
\begin{equation}
\left( \frac{\partial }{\partial t}+\mathbf{v}\cdot \nabla \right) f(\mathbf{%
r},\mathbf{v},t)=J[\mathbf{r},\mathbf{v},t|f]  \label{x1}
\end{equation}%
where the collision operator $J[\mathbf{r},\mathbf{v},t|f]$ is a function of
position and velocity and a \emph{functional} of the distribution function.
No particular details of the form of the collision operator will be
important here but all results are formulated with the examples of BGK-type
relaxation models, the Boltzmann equation and the Enskog equation in mind.
The first five velocity moments of $f$ define the number density 
\begin{equation}
n(\mathbf{r},t)=\int \;d\mathbf{v}f(\mathbf{r},\mathbf{v},t),
\label{density}
\end{equation}%
the flow velocity 
\begin{equation}
\mathbf{u}(\mathbf{r},t)=\frac{1}{n(\mathbf{r},t)}\int \;d\mathbf{v}\mathbf{v%
}f(\mathbf{r},\mathbf{v},t),  \label{velocity}
\end{equation}%
and the kinetic temperature 
\begin{equation}
T(\mathbf{r},t)=\frac{m}{Dn(\mathbf{r},t)k_{B}}\int \;d\mathbf{v}C^{2}(%
\mathbf{r},t)f(\mathbf{r},\mathbf{v},t),  \label{temperature}
\end{equation}%
where $\mathbf{C}(\mathbf{r},t)\equiv \mathbf{v}-\mathbf{u}(\mathbf{r},t)$
is the peculiar velocity. The macroscopic balance equations for density $n$,
momentum $m\mathbf{u}$, and energy $\frac{D}{2}nk_{B}T$ follow directly from
eq.\ ({\ref{x1}) by multiplying with $1$, $m\mathbf{v}$, and $\frac{1}{2}%
mC^{2}$ and integrating over $\mathbf{v}$: 
\begin{eqnarray}
D_{t}n+n\nabla \cdot \mathbf{u} &=&0\;  \label{x2} \\
D_{t}u_{i}+(mn)^{-1}\nabla _{j}P_{ij} &=&0  \notag \\
D_{t}T+\frac{2}{Dnk_{B}}\left( \nabla \cdot \mathbf{q}+P_{ij}\nabla
_{j}u_{i}\right)  &=&-\zeta T,  \notag
\end{eqnarray}%
where $D_{t}=\partial _{t}+\mathbf{u}\cdot \nabla $ is the convective
derivative. The microscopic expressions for the pressure tensor $\mathsf{P=P}%
\left[ f\right] $, the heat flux $\mathbf{q=q}\left[ f\right] $} depend on
the exact form of the collision operator (see ref. \cite{McLennan},\cite%
{LutskoJCP} for a general discussion){\ but as indicated, they are in
general functionals of the distribution, while the cooling rate $\zeta $ is
given by 
\begin{equation}
\zeta (\mathbf{r},t)=\frac{1}{Dn(\mathbf{r},t)k_{B}T(\mathbf{r},t)}\int \,d%
\mathbf{v}mC^{2}J[\mathbf{r},\mathbf{v},t|f].  \label{heating}
\end{equation}%
}

\subsection{Formulation of the gradient expansion}

The goal of the Chapman-Enskog method is to construct a so-called \emph{%
normal }solution to the kinetic equation, eq.(\ref{x1}). In the standard
formulation of the method\cite{McLennan}, this is defined as a distribution $%
f(\mathbf{r},\mathbf{v},t)$ for which all of the space and time dependence
occurs through the hydrodynamic variables, denoted collectively as $\psi
\equiv \left\{ n,\mathbf{u},T\right\} $, and their derivatives so that%
\begin{equation}
f(\mathbf{r},\mathbf{v},t)=f\left( \mathbf{v};\psi \left( \mathbf{r}%
,t\right) ,\mathbf{\nabla }\psi \left( \mathbf{r},t\right) ,\mathbf{\nabla
\nabla }\psi \left( \mathbf{r},t\right) ,...\right) .  \label{KE}
\end{equation}%
The distribution is therefore a \emph{functional} of the fields $\psi \left( 
\mathbf{r},t\right) $ or, equivalently in this case, a \emph{function} of
the fields and their gradients to all orders. In the following, this
particular type of functional dependence will be denoted more compactly with
the notation $f\left( \mathbf{v};\left[ \mathbf{\nabla }^{(n)}\psi \left( 
\mathbf{r},t\right) \right] \right) $ where the index, $n$, indicates the
maximum derivative that is used. When all derivatives are possible, as in
eq.(\ref{KE}) the notation $f(\mathbf{r},\mathbf{v},t)=f\left( \mathbf{v};%
\left[ \mathbf{\nabla }^{(\infty )}\psi \left( \mathbf{r},t\right) \right]
\right) $ will be used. The kinetic equation, eq.(\ref{x1}), the balance
equations, eqs.(\ref{x2}), and the definitions of the various fluxes and
sources then provide a closed set of equations from which to determine the
distribution. Note that since the fluxes and sources are functionals of the
distribution, their space and time dependence also occurs implicitly via
their dependence on the hydrodynamic fields and their derivatives.

Given such a solution has been found for a particular set of boundary
conditions yielding the hydrodynamic state $\psi _{0}\left( \mathbf{r}%
,t\right) $ with distribution $f_{0}\left( \mathbf{v};\left[ \mathbf{\nabla }%
^{(\infty )}\psi _{0}\left( \mathbf{r},t\right) \right] \right) $, the aim
is to describe deviations about this state, denoted $\delta \psi $, so that
the total hydrodynamic fields are $\psi =\psi _{0}+\delta \psi $. In the
Chapman-Enskog method, it is assumed that the deviations are smooth in the
sense that 
\begin{equation}
\delta \psi \gg l\mathbf{\nabla }\delta \psi \gg l^{2}\mathbf{\nabla \nabla }%
\delta \psi ...,
\end{equation}%
where $l$ is the mean free path, so that one can work perturbatively in
terms of the gradients of the perturbations to the hydrodynamic fields. To
develop this perturbation theory systematically, it is convenient to
introduce a fictitious small parameter, $\epsilon $, and to write the
gradient operator as $\mathbf{\nabla }=\mathbf{\nabla }^{\left( 0\right)
}+\epsilon \mathbf{\nabla }^{\left( 1\right) }$ where the two operators on
the right are defined by $\mathbf{\nabla }_{0}\psi =\mathbf{\nabla }\psi _{0}
$ and $\mathbf{\nabla }_{1}\psi =\mathbf{\nabla \delta }\psi $. This then
generates an expansion of the distribution that looks like%
\begin{eqnarray}
f\left( \mathbf{v};\left[ \mathbf{\nabla }^{(\infty )}\psi \left( \mathbf{r}%
,t\right) \right] \right)  &=&f^{(0)}\left( \mathbf{v};\mathbf{\nabla }%
_{0}^{\left( \infty \right) }\psi \left( \mathbf{r},t\right) \right) 
\label{dist-expansion} \\
&&+\epsilon f^{(1)}\left( \mathbf{v};\mathbf{\nabla }_{1}\mathbf{\delta }%
\psi ,\mathbf{\nabla }_{0}^{\left( \infty \right) }\psi \left( \mathbf{r}%
,t\right) \right)   \notag \\
&&+\epsilon ^{2}f^{(2)}\left( \mathbf{v};\mathbf{\nabla }_{1}\mathbf{\nabla }%
_{1}\mathbf{\delta }\psi ,\left( \mathbf{\nabla }_{1}\mathbf{\delta }\psi
\right) ^{2},\mathbf{\nabla }_{0}^{(\infty )}\psi \left( \mathbf{r},t\right)
\right)   \notag \\
&&+...  \notag
\end{eqnarray}%
where $f^{(1)}$ will be linear in $\mathbf{\nabla }_{1}\mathbf{\delta }\psi $%
, $f^{(2)}$ will be linear in $\mathbf{\nabla }_{1}\mathbf{\nabla }_{1}%
\mathbf{\delta }\psi $ and $\left( \mathbf{\nabla }_{1}\mathbf{\delta }\psi
\right) ^{2}$, etc. This notation is meant to be taken literally:\ the
quantity $\mathbf{\nabla }_{0}^{(\infty )}\psi \left( \mathbf{r},t\right)
=\left\{ \psi \left( \mathbf{r},t\right) ,\mathbf{\nabla }_{0}\psi \left( 
\mathbf{r},t\right) ,...\right\} =\left\{ \psi \left( \mathbf{r},t\right) ,%
\mathbf{\nabla }\psi _{0}\left( \mathbf{r},t\right) ,...\right\} $ so that
at each order in perturbation theory, the distribution is a function of the
exact field $\psi \left( \mathbf{r},t\right) $ as well as all gradients of
the reference field. This involves a departure from the usual formulation of
the Chapman-Enskog definition of a normal state. In the standard form, the
distribution is assumed to be a functional of the \emph{exact} fields $\psi
\left( \mathbf{r},t\right) $ whereas here it is  proposed that the
distribution is a functional of the exact field $\psi \left( \mathbf{r}%
,t\right) $ \emph{and} of the reference state $\psi _{0}\left( \mathbf{r}%
,t\right) $. Of course, it is obvious that in order to study deviations
about a reference state within the Chapman-Enskog framework, the
distribution will have to be a functional of that reference state.
Nevertheless, this violates, or generalizes, the usual definition of a
normal solution since there are now two sources of space and time dependence
in the distribution:\ the exact hydrodynamics fields and the reference
hydrodynamic state. For deviations from an equilibrium state, this point is
moot since $\mathbf{\nabla }\psi _{0}\left( \mathbf{r},t\right) =0$, etc.

The perturbative expansion of the distribution will generate a similar
expansion of the fluxes and sources through their functional dependence on
the distribution, see e.g. eq.(\ref{heating}), so that one writes%
\begin{equation}
P_{ij}=P_{ij}^{(0)}+\epsilon P_{ij}^{(1)}+...
\end{equation}%
and so forth. Since the balance equations link space and time derivatives,
it is necessary to introduce a multiscale expansion of the time derivatives
in both the kinetic equation and the balance equations as 
\begin{equation}
\frac{\partial }{\partial t}f=\partial _{t}^{(0)}f+\epsilon \partial
_{t}^{(1)}f+...
\end{equation}%
The precise meaning of the symbols $\partial _{t}^{(0)}$, $\partial
_{t}^{(1)}$ is that the balance equations define $\partial _{t}^{(i)}$ in
terms of the spatial gradients of the hydrodynamic fields and these
definitions, together with the normal form of the distribution, define the
action of these symbols on the distribution. Finally, to maintain
generality, note that sometimes (specifically in the Enskog theory) the
collision operator itself is non-local and must be expanded as well in
gradients in $\delta \psi $ so that we write%
\begin{equation}
J[\mathbf{r},\mathbf{v},t|f]=J_{0}[\mathbf{r},\mathbf{v},t|f]+\epsilon J_{1}[%
\mathbf{r},\mathbf{v},t|f]+...
\end{equation}%
and it is understood that $J_{0}[\mathbf{r},\mathbf{v},t|f]$ by definition
involves no gradients with respect to the perturbations $\delta \psi \left( 
\mathbf{r},t\right) $ but will, in general, contain gradients of \emph{all}
orders in the reference fields $\psi _{0}\left( \mathbf{r},t\right) $. (Note
that the existence of a normal solution is plausible if the spatial and
temporal dependence of the collision operator is also normal which is, in
fact, generally the case. However, for simplicity, no effort is made here to
indicate this explicitly.) A final property of the perturbative expansion
concerns the relation between the various distributions and the hydrodynamic
variables. The zeroth order distribution is required to reproduce the exact
hydrodynamic variables via%
\begin{equation}
\left( 
\begin{array}{c}
n(\mathbf{r},t) \\ 
n(\mathbf{r},t)\mathbf{u}(\mathbf{r},t) \\ 
Dn(\mathbf{r},t)k_{B}T%
\end{array}%
\right) =\int \left( 
\begin{array}{c}
1 \\ 
\mathbf{v} \\ 
mC^{2}%
\end{array}%
\right) f^{(0)}\left( \mathbf{v};\mathbf{\nabla }_{0}^{\left( \infty \right)
}\psi \left( \mathbf{r},t\right) \right) d\mathbf{v}  \label{f0-hydro}
\end{equation}%
while the higher order terms are orthogonal to the first three velocity
moments%
\begin{equation}
\int \left( 
\begin{array}{c}
1 \\ 
\mathbf{v} \\ 
mC^{2}%
\end{array}%
\right) f^{(n)}\left( \mathbf{v};\mathbf{\nabla }_{0}^{\left( \infty \right)
}\psi \left( \mathbf{r},t\right) \right) d\mathbf{v}=0,\;n>0,
\label{fn-hydro}
\end{equation}%
so that the total distribution $f=f^{(0)}+f^{(1)}+...$ satisfies eqs.(\ref%
{density})-(\ref{temperature}).

\subsection{The reference state}

Recall that the goal is to describe deviations from the reference state $%
\psi _{0}\left( \mathbf{r},t\right) $ which corresponds to the distribution $%
f_{0}\left( \mathbf{r},\mathbf{v,}t;\left[ \psi _{0}\right] \right) $ and in
fact the distribution and fields are related by the definitions given in
eqs.(\ref{density})-(\ref{temperature}). The reference distribution is
itself assumed to be normal so that the dependence on $\mathbf{r}$ and $t$
occurs implicitly through the fields. In terms of the notation used here,
the reference distribution satisfies the kinetic equation, eq.(\ref{KE}),
and the full, nonlinear balance equations, eqs.(\ref{x2}). Using the
definitions given above, these translate into 
\begin{equation}
\left( \partial _{t}^{\left( 0\right) }+\mathbf{v}\cdot \nabla ^{\left(
0\right) }\right) f_{0}\left( \mathbf{r},\mathbf{v,}t;\left[ \psi _{0}\right]
\right) =J_{0}[\mathbf{r},\mathbf{v},t|f_{0}]  \label{ref-KE}
\end{equation}%
and the fields are solutions to the full, nonlinear balance equations {%
\begin{eqnarray}
\partial _{t}^{\left( 0\right) }{n}_{0}{+\mathbf{u}\cdot }\mathbf{\nabla }%
^{\left( 0\right) }n_{0}+n_{0}\mathbf{\nabla }^{\left( 0\right) }\cdot 
\mathbf{u}_{0} &=&0\;  \label{ref-balance} \\
\partial _{t}^{\left( 0\right) }{u}_{0,i}{+\mathbf{u}}_{0}{\cdot }\mathbf{%
\nabla }^{\left( 0\right) }u_{0.i}+(mn_{0})^{-1}\partial
_{j}^{(0)}P_{ij}^{(00)} &=&0  \notag \\
{\partial _{t}^{(0)}T}_{0}{+\mathbf{u}}_{0}{\cdot }\mathbf{\nabla }^{\left(
0\right) }T_{0}+\frac{2}{Dn_{0}k_{B}}\left( \mathbf{\nabla }^{\left(
0\right) }\cdot \mathbf{q}^{(00)}+P_{ij}^{(00)}\partial
_{j}^{(0)}u_{0,i}\right)  &=&-\zeta ^{(00)}T_{0}\;,  \notag
\end{eqnarray}%
where, e.g., }$P_{ij}^{(00)}$ is the pressure tensor evaluated in the
reference state, and 
\begin{equation}
{\partial _{t}^{(n)}\psi }_{0}=0,\;n>0.
\end{equation}%
Thus, in the ordering scheme developed here, the reference state is an exact
solution to the zeroth order perturbative equations.

For the standard case describing deviations from the equilibrium state, the
hydrodynamic fields are constant in both space and time and $\zeta ^{(00)}=0$
so that the balance equations just reduce to ${\partial _{t}^{(0)}\psi }%
_{0}=0$. The left hand side of the kinetic equation therefore vanishes
leaving $0=J_{0}[\mathbf{r},\mathbf{v},t|f_{0}]$ which is indeed satisfied
by the equilibrium distribution. For a granular fluid, $\zeta ^{(00)}\neq 0$
and the simplest solution that can be constructed consists of spatially
homogeneous, but time dependent fields giving%
\begin{equation}
\partial _{t}^{\left( 0\right) }f_{0}\left( \mathbf{r},\mathbf{v,}t;\left[
\psi _{0}\right] \right) =J_{0}[\mathbf{r},\mathbf{v},t|f_{0}]  \label{e1}
\end{equation}%
and {%
\begin{eqnarray}
\partial _{t}^{\left( 0\right) }{n}_{0} &=&0\; \\
\partial _{t}^{\left( 0\right) }{u}_{0,i} &=&0  \notag \\
{\partial _{t}^{(0)}T}_{0} &=&-\zeta ^{(00)}T_{0}  \notag
\end{eqnarray}%
so that the distribution depends on time through its dependence on the
temperature. The balance equations, together with the assumption of
normality, serve to define the meaning of the left hand side of eq.(\ref{e1}%
) giving}%
\begin{equation}
-\zeta ^{(00)}T_{0}\frac{\partial }{\partial T}f_{0}\left( \mathbf{r},%
\mathbf{v,}t;\left[ \psi _{0}\right] \right) =J_{0}[\mathbf{r},\mathbf{v}%
,t|f_{0}].
\end{equation}%
Typically, this is solved by assuming a scaling solution of the form {\ }$%
f_{0}\left( \mathbf{r},\mathbf{v,}t;\left[ \psi _{0}\right] \right) =\Phi
\left( \mathbf{v}\sqrt{\frac{m\sigma ^{2}}{k_{B}T\left( t\right) }}\right) $.

\subsection{The zeroth order Chapman-Enskog solution}

As emphasized above, the Chapman-Enskog method is an expansions in gradients
of the deviations of the hydrodynamic fields from the reference state. Using
the ordering developed above,{the zeroth order kinetic equation is}%
\begin{equation}
\partial _{t}^{(0)}f^{(0)}\left( \mathbf{r},\mathbf{v};\delta \psi \left( 
\mathbf{r},t\right) ,\left[ \psi _{0}\right] \right) +\mathbf{v}\cdot \nabla
^{(0)}f^{(0)}\left( \mathbf{r},\mathbf{v};\delta \psi \left( \mathbf{r}%
,t\right) ,\left[ \psi _{0}\right] \right) =J_{0}[\mathbf{r},\mathbf{v}%
,t|f_{0}].  \label{zero-KE}
\end{equation}%
and the zeroth order balance equations are{%
\begin{eqnarray}
{\partial _{t}^{(0)}n+\mathbf{u}\cdot }\mathbf{\nabla }n_{0}+n\mathbf{\nabla 
}\cdot \mathbf{u}_{0} &=&0\;  \label{zero-KE-2} \\
{\partial _{t}^{(0)}u}_{i}{+\mathbf{u}\cdot }\mathbf{\nabla }%
u_{0.i}+(mn)_{j}^{-1}\mathbf{\nabla }^{\left( 0\right) }P_{ij}^{(0)} &=&0 
\notag \\
{\partial _{t}^{(0)}T+\mathbf{u}\cdot \nabla }T_{0}+\frac{2}{Dnk_{B}}\left( 
\mathbf{\nabla }^{\left( 0\right) }\cdot \mathbf{q}^{(0)}+P_{ij}^{(0)}%
\partial _{j}u_{0,i}\right)  &=&-\zeta ^{(0)}T.  \notag
\end{eqnarray}%
Making use of the balance equations satisfied by the reference fields, (\ref%
{ref-balance}), this can be written in terms of the deviations as 
\begin{eqnarray}
{\partial _{t}^{(0)}\delta n+\delta \mathbf{u}\cdot \nabla }n_{0}+\delta
n\nabla \cdot \mathbf{u}_{0} &=&0\;  \label{zero-balance} \\
{\partial _{t}^{(0)}\delta u}_{i}{+\delta \mathbf{u}\cdot \nabla }%
u_{0.i}+(mn)^{-1}\nabla _{j}^{(0)}P_{ij}^{(0)}-(mn_{0})^{-1}\nabla
_{j}P_{ij}^{(00)} &=&0  \notag \\
{\partial _{t}^{(0)}\delta T+\delta \mathbf{u}\cdot \nabla }T_{0}+\frac{2}{%
Dnk_{B}}\left( \nabla ^{(0)}\cdot \mathbf{q}^{(0)}+P_{ij}^{(0)}\nabla
_{j}u_{0,i}\right) -\frac{2}{Dn_{0}k_{B}}\left( \nabla \cdot \mathbf{q}%
^{(00)}+P_{ij}^{(00)}\nabla _{j}u_{0,i}\right)  &=&-\zeta ^{(0)}T+\zeta
^{(00)}T_{0}.  \notag
\end{eqnarray}%
Since the zeroth-order distribution is a \emph{function} of }$\delta \psi $
but a \emph{functional} of the reference fields, t{he time derivative in eq.(%
\ref{zero-KE}) is evaluated using }%
\begin{equation}
\partial _{t}^{(0)}f^{(0)}=\sum_{\alpha }\left( {\partial _{t}^{(0)}}\delta
\psi _{\alpha }\left( \mathbf{r},t\right) \right) \frac{\partial }{\partial
\delta \psi _{\alpha }\left( \mathbf{r},t\right) }f^{(0)}+\sum_{\alpha }\int
d\mathbf{r}^{\prime }\;\left( {\partial _{t}^{(0)}}\psi _{0,\alpha }\left( 
\mathbf{r}^{\prime },t\right) \right) \frac{\delta }{\delta \psi _{0,\alpha
}\left( \mathbf{r}^{\prime },t\right) }f^{(0)}.  \label{zero-t}
\end{equation}%
and these equations must be solved subject to the additional boundary
condition%
\begin{equation}
\lim_{\delta \psi \rightarrow 0}f^{(0)}\left( \mathbf{r},\mathbf{v},t;\delta
\psi \left( \mathbf{r},t\right) ,\left[ \psi _{0}\right] \right)
=f_{0}\left( \mathbf{r},\mathbf{v},t;\left[ \psi _{0}\right] \right) .
\label{bc0}
\end{equation}%
There are several important points to be made here. First, it must be
emphasized that the reference fields $\psi _{0}\left( \mathbf{r},t\right) $
and the deviations $\delta \psi \left( \mathbf{r},t\right) $ are playing
different roles in these equations. The former are fixed and assumed known
whereas the latter are independent variables. The result of a solution of
these equations will be the zeroth order distribution as a function of the
variables $\delta \psi $. For any given physical problem, the deviations
will be determined by solving the balance equations, eqs.(\ref{zero-balance}%
), subject to appropriate boundary conditions and only then is the
distribution completely specified. Second, nothing is said here about the
solution of eqs.(\ref{zero-KE})-(\ref{zero-t}) which, in general, constitute
a complicated functional equation in terms of the reference state variables $%
\psi _{0,\alpha }\left( \mathbf{r},t\right) $ . The only obvious exceptions,
and perhaps the only practical cases, are when the reference state is either
time-independent, so that ${\partial _{t}^{(0)}}\psi _{0,\alpha }=0$, or
spatially homogeneous so that $f^{(0)}$is a function, and not a functional,
of the reference fields. The equilibrium state is both, the homogeneous
cooling state is a spatially homogeneous state and time-independent flow
states such as uniform shear flow or Pouseille flow with thermalizing walls
are important examples of time-independent, spatially inhomogeneous states.
Third, since eqs.(\ref{zero-KE})-(\ref{zero-KE-2}) are the lowest order
equations in a gradient expansion, they are to be solved for \emph{%
arbitrarily large} deviations of the fields, $\delta \psi $. There is no
sense in which the deviations should be considered to be small. The fourth
observation, and perhaps the most important, is that there is no conceptual
connection between the zeroth order distribution $f^{(0)}\left( \mathbf{v}%
;\delta \psi \left( \mathbf{r},t\right) ,\mathbf{\nabla }_{0}^{(\infty
)}\psi _{0}\left( \mathbf{r},t\right) \right) $ and the reference
distribution $f_{0}\left( \mathbf{v};\mathbf{\nabla }_{0}^{(\infty )}\psi
_{0}\left( \mathbf{r},t\right) \right) $ except for the limit given in eq.(%
\ref{bc0}). In particular, it will almost always be the case that 
\begin{equation}
f^{(0)}\left( \mathbf{v};\delta \psi \left( \mathbf{r},t\right) ,\mathbf{%
\nabla }_{0}^{(\infty )}\psi _{0}\left( \mathbf{r},t\right) \right) \neq
f_{0}\left( \mathbf{v};\mathbf{\nabla }_{0}^{(\infty )}\left( \psi
_{0}\left( \mathbf{r},t\right) +\delta \psi \left( \mathbf{r},t\right)
\right) \right) .
\end{equation}%
A rare exception for which this inequality is reversed is when the reference
state is the equilibrium state. In that case, the density, temperature and
velocity fields are uniform and the reference distribution is just a Gaussian%
\begin{equation}
f_{0}\left( \mathbf{r},\mathbf{v};\mathbf{\nabla }_{0}^{(\infty )}\psi
_{0}\right) =\phi _{0}\left( \mathbf{v};n_{0},T_{0},\mathbf{U}_{0}\right) 
\end{equation}%
and the solution to the zeroth order equations is the local equilibrium
distribution 
\begin{equation}
f^{(0)}\left( \mathbf{v};\delta \psi \left( \mathbf{r},t\right) ,\mathbf{%
\nabla }_{0}^{(\infty )}\psi _{0}\left( \mathbf{r},t\right) \right) =\phi
_{0}\left( \mathbf{v};n+\delta n\left( \mathbf{r},\mathbf{t}\right)
,T+\delta T\left( \mathbf{r},\mathbf{t}\right) ,\mathbf{U}+\delta \mathbf{U}%
\left( \mathbf{r},\mathbf{t}\right) \right) =f_{0}\left( \mathbf{v};\mathbf{%
\nabla }_{0}^{(\infty )}\left( \psi _{0}\left( \mathbf{r},t\right) +\delta
\psi \left( \mathbf{r},t\right) \right) \right) .  \label{localize}
\end{equation}%
For steady states, as will be illustrated in the next Section, it is not the
case that $f^{(0)}$ is obtained from the steady-state distribution via a
''localization'' along the lines of that shown in eq.(\ref{localize}). On
the other hand, eqs.(\ref{zero-KE})-(\ref{zero-KE-2}) are the same whether
they are solved for the general field $\delta \psi \left( \mathbf{r}%
,t\right) \;$or for the spatially homogeneous field $\delta \psi \left(
t\right) $ with the subsequent localization $\delta \psi \left( t\right)
\rightarrow \delta \psi \left( \mathbf{r},t\right) $. Furthermore, these
equations are identical to those one would solve in order to obtain an exact
normal solution to the full kinetic equation, eq.(\ref{ref-KE}) and balance
equations, eq.(\ref{ref-balance}), for the fields $\psi _{0}\left( \mathbf{r}%
,t\right) +\delta \psi \left( t\right) $. In other words, the zeroth-order
Chapman-Enskog distribution is the localization of the exact distribution
for homogeneous deviations from the reference state. Again, only in the case
of the equilibrium reference state is it true that this corresponds to the
localization of the reference state itself.

\subsection{First order Chapman-Enskog}

In the following, the equations for the first-order terms will also be
needed. Collecting terms in eq.(\ref{ref-KE}), the first order distribution
function is found to satisfy%
\begin{eqnarray}
&&\partial _{t}^{(0)}f^{(1)}(\mathbf{v};\delta \psi \left( \mathbf{r}%
,t\right) ,\left[ \psi _{0}\right] )+\mathbf{v}\cdot \mathbf{\nabla }%
^{(0)}f^{(1)}(\mathbf{v};\delta \psi \left( \mathbf{r},t\right) ,\left[ \psi
_{0}\right] )  \label{f1} \\
&=&J_{0}[\mathbf{r},\mathbf{v},t|f_{1}]+J_{1}[\mathbf{r},\mathbf{v}%
,t|f_{0}]-\left( \partial _{t}^{(1)}f^{(0)}(\mathbf{v};\delta \psi \left( 
\mathbf{r},t\right) ,\left[ \psi _{0}\right] )+\mathbf{v}\cdot \nabla
^{(1)}f^{(0)}(\mathbf{v};\delta \psi \left( \mathbf{r},t\right) ,\left[ \psi
_{0}\right] )\right)   \notag
\end{eqnarray}%
and the first-order balance equations become{%
\begin{eqnarray}
{\partial _{t}^{(1)}\delta n+\mathbf{u}\cdot \nabla }\delta n+n\nabla \cdot
\delta \mathbf{u} &=&0\;  \label{P1} \\
{\partial _{t}^{(1)}\delta u_{i}+\mathbf{u}\cdot }\mathbf{\nabla }{\delta }%
u_{i}+(mn)^{-1}\nabla _{j}^{\left( 1\right) }P_{ij}^{\left( 0\right)
}+(mn)^{-1}\nabla _{j}^{\left( 0\right) }P_{ij}^{\left( 1\right) } &=&0 
\notag \\
{\partial _{t}^{(1)}\delta T+\mathbf{u}\cdot \mathbf{\nabla }\delta }T+\frac{%
2}{Dnk_{B}}\left( \mathbf{\nabla }^{\left( 1\right) }\cdot \mathbf{q}%
^{(0)}+P_{ij}^{\left( 0\right) }\nabla _{j}\delta u_{i}\right) +\frac{2}{%
Dnk_{B}}\left( \mathbf{\nabla }^{\left( 0\right) }\cdot \mathbf{q}%
^{(1)}+P_{ij}^{\left( 1\right) }\nabla _{j}u_{0,i}\right)  &=&-\zeta ^{(1)}T.
\notag
\end{eqnarray}%
}

\section{Application to Uniform Shear Flow of Granular Fluids}

Uniform shear flow (USF) is a macroscopic state that is characterized by a
constant density, a uniform temperature and a simple shear with the local
velocity field given by 
\begin{equation}
u_{i}=a_{ij}r_{j},\quad a_{ij}=a\delta _{ix}\delta _{jy},  \label{profile}
\end{equation}%
where $a$ is the \emph{constant} shear rate. If one assumes that the
pressure tensor, heat flux vector and cooling rate are also spatially
uniform, the reference-state balance equations, eqs.(\ref{ref-balance}),
become{%
\begin{eqnarray}
\partial _{t}^{\left( 0\right) }{n}_{0} &=&0\;  \label{ssx} \\
\partial _{t}^{\left( 0\right) }{u}_{0,i}+au_{0,y}\delta _{ix} &=&0  \notag
\\
{\partial _{t}^{(0)}T}_{0}+\frac{2}{Dn_{0}k_{B}}aP_{xy}^{(00)} &=&-\zeta
^{(00)}T_{0}\;,  \notag
\end{eqnarray}%
The question of whether or not these assumptions of spatial homogeneity are
true depends on the detailed form of the collision operator:\ in ref.\cite%
{LutskoPolydisperse} it is shown that only for the linear velocity profile,
eq.(\ref{profile}), this assumption }is easily verified for the Enskog
kinetic theory (and hence for simpler approximations to it such as the
Boltzmann and BGK theories).{\ }This linear velocity profile is generated by
Lee-Edwards boundary conditions \cite{LeesEdwards}, which are simply
periodic boundary conditions in the local Lagrangian frame. For elastic
gases, $\zeta ^{(00)}=0$ and the temperature grows in time due to viscous
heating and so a steady state is not possible unless an external
(artificial) thermostat is introduced\cite{MirimThesisArticle}. However, for
inelastic gases, the temperature changes in time due to the competition
between two (opposite) mechanisms: on the one hand, viscous (shear) heating
and, on the other hand, energy dissipation in collisions. A steady state
occurs when both mechanisms cancel each other at which point the balance
equation for temperature becomes 
\begin{equation}
\frac{2}{Dn_{0}k_{B}}aP_{xy}^{(00)}=-\zeta ^{(00)}T_{0}.
\end{equation}%
Note that both the pressure tensor and the cooling rate are in general
functions of the two control parameters, the shear rate and the coefficient
of restitution, and the hydrodynamic variables, the density and the
temperature, so that this relation fixes any one of these in terms of the
the other three: for example, it could be viewed as giving the steady-state
temperature as a function of the other variables.

At a microscopic level, the one-body distribution for USF will clearly be
inhomogeneous since the eq.(\ref{velocity}) and eq.(\ref{profile}) imply
that the steady-state distribution must give%
\begin{equation}
ay\widehat{\mathbf{x}}=\frac{1}{n_{0}}\int \;d\mathbf{v}\mathbf{v}f_{0}(%
\mathbf{r},\mathbf{v}).
\end{equation}%
However, it can be shown, at least up to the Enskog theory{\cite%
{LutskoPolydisperse}}, that for the Lees-Edwards boundary conditions, the
state of USF possesses a modified translational invariance whereby the
steady state distribution, when expressed in terms of the local rest-frame
velocities $V_{i}=v_{i}-a_{ij}r_{j}$ does not have any explicit dependence
on position. In terms of these variables, and assuming a steady state, the
kinetic equation becomes 
\begin{equation}
-aV_{y}\frac{\partial }{\partial V_{x}}f(\mathbf{V})=J\left[ \mathbf{V}|f,f%
\right] \;.  \label{2.15}
\end{equation}%
The solution of this equation has been considered in some detail for the
BGK-type models\cite{MirimThesis},\cite{MirimThesisArticle},\cite%
{Brey_EarlyKineticModels},\cite{Brey_KineticModels}, the Boltzmann equation%
\cite{SelGoldhirsch}, and the Enskog equation\cite{ChouRichman1},\cite%
{ChouRichman2},\cite{LutskoPolydisperse}.\textbf{\ }

\subsection{The model kinetic theory}

Here, for simplicity, attention will be restricted to a particularly simple
kinetic theory which nevertheless gives realistic results that can be
compared to experiment. The kinetic theory used is the kinetic model of
Brey, Dufty and Santos\cite{Brey_KineticModels}, which is a relaxation type
model where the operator $J[f,f]$ is approximated as 
\begin{equation}
J[f,f]\rightarrow -\nu ^{\ast }\left( \alpha \right) \nu \left( \psi \right)
(f-\phi _{0})+\frac{1}{2}\zeta ^{\ast }\left( \alpha \right) \nu \left( \psi
\right) \frac{\partial }{\partial \mathbf{v}}\cdot \left( \mathbf{C}f\right)
.  \label{BGK}
\end{equation}%
The right hand side involves the peculiar velocity $\mathbf{C}=\mathbf{v}-%
\mathbf{u}=\mathbf{V}-\delta \mathbf{u}$ and the local equilibrium
distribution, eq.(\ref{le}). The parameters in this relaxation approximation
are taken so as to give agreement with the results from the Boltzmann theory
of the homogeneous cooling state as discussed in ref.\cite%
{Brey_KineticModels}. Defining the collision rate for elastic hard spheres
in the Boltzmann approximation as 
\begin{equation}
\nu \left( \psi \right) =\frac{8\pi ^{\left( D-2\right) /2}}{\left(
D+2\right) \Gamma \left( D/2\right) }n\sigma ^{D}\sqrt{\frac{\pi k_{B}T}{%
m\sigma ^{2}}},
\end{equation}%
the correction for the effect of the inelasticity is chosen to reproduce the
Navier-Stokes shear viscosity coefficient of an inelastic gas of hard
spheres in the Boltzmann approximation\cite{BreyCubero},\cite{LutskoCE}
giving 
\begin{equation}
\nu ^{\ast }\left( \alpha \right) =\frac{1}{4D}\left( 1+\alpha \right)
\left( \left( D-1\right) \alpha +D+1\right) .
\end{equation}%
The second term in eq.(\ref{BGK}) accounts for the collisional cooling and
the coefficient is chosen so as to give the same cooling rate for the
homogeneous cooling state as the Boltzmann kinetic theory\cite{BreyCubero},%
\cite{LutskoCE}, 
\begin{equation}
\zeta ^{\ast }\left( \alpha \right) =\frac{D+2}{4D}\left( 1-\alpha
^{2}\right) .
\end{equation}%
In this case, the expressions for the pressure tensor, heat-flux vector and
cooling rate take particularly simple forms typical of the Boltzmann
description\cite{ChapmanCowling}%
\begin{eqnarray}
P_{ij} &=&m\int d\mathbf{C}\;C_{i}C_{j}f\left( \mathbf{r,C,}t\right) ,
\label{fluxBGK} \\
q_{i} &=&\frac{1}{2}m\int d\mathbf{C}\;C_{i}C^{2}f\left( \mathbf{r,C,}%
t\right) ,  \notag
\end{eqnarray}%
while the cooling rate can be calculated directly from eqs.(\ref{BGK}) and (%
\ref{heating}) with the result $\zeta (\psi )=\nu \left( \psi \right) \zeta
^{\ast }\left( \alpha \right) $.

\subsection{The steady-state}

Before proceeding with the Chapman-Enskog solution of the kinetic equation,
it is useful to describe the steady state for which the distribution
satisfies eq.(\ref{2.15}) which now becomes%
\begin{equation}
-aV_{y}\frac{\partial }{\partial V_{x}}f(\mathbf{V})=-\nu ^{\ast }\left(
\alpha \right) \nu \left( \psi _{0}\right) (f-\phi _{0})+\frac{1}{2}\zeta
^{\ast }\left( \alpha \right) \nu \left( \psi _{0}\right) \frac{\partial }{%
\partial \mathbf{V}}\cdot \left( \mathbf{V}f\right) .  \label{kss}
\end{equation}%
The balance equations reduce to 
\begin{equation}
2aP_{xy}^{ss}=-\zeta ^{\ast }\left( \alpha \right) \nu \left( \psi \right)
Dn_{0}k_{B}T_{0}.  \label{ss}
\end{equation}%
An equation for the pressure tensor is obtained by multiplying eq.(\ref{kss}%
) through by $mV_{i}V_{j}$ and integrating giving%
\begin{equation*}
aP_{iy}^{ss}\delta _{jx}+aP_{jy}^{ss}\delta _{ix}=-\nu ^{\ast }\left( \alpha
\right) \nu \left( \psi _{0}\right) (P_{ij}^{ss}-n_{0}k_{B}T_{0}\delta
_{ij})-\zeta ^{\ast }\left( \alpha \right) \nu \left( \psi _{0}\right)
P_{ij}^{ss}.
\end{equation*}%
This set of algebraic equations is easily solved giving the only non-zero
components of the pressure tensor as 
\begin{eqnarray}
P_{ii}^{ss} &=&\frac{\nu ^{\ast }\left( \alpha \right) +\delta _{ix}D\zeta
^{\ast }\left( \alpha \right) }{\nu ^{\ast }\left( \alpha \right) +\zeta
^{\ast }\left( \alpha \right) }n_{0}k_{B}T_{0} \\
P_{xy}^{ss} &=&-\frac{a_{ss}^{\ast }}{\nu ^{\ast }\left( \alpha \right)
+\zeta ^{\ast }\left( \alpha \right) }P_{yy},  \notag
\end{eqnarray}%
where $a_{ss}^{\ast }=a_{ss}/\nu \left( \psi _{0}\right) $ satisfies the
steady-state condition, eq.(\ref{ss})%
\begin{equation}
\frac{a_{ss}^{\ast 2}\nu ^{\ast }\left( \alpha \right) }{\left( \nu ^{\ast
}\left( \alpha \right) +\zeta ^{\ast }\left( \alpha \right) \right) ^{2}}=%
\frac{D}{2}\zeta ^{\ast }\left( \alpha \right) .  \label{balance}
\end{equation}%
For fixed control parameters, $\alpha $ and $a$, this is a relation
constraining the state variables $n_{0}$ and $T_{0}$. The steady-state
distribution can be given explicitly, see e.g. \cite{SantosSolveBGK}.

\subsection{Zeroth order Chapman-Enskog}

Since the only spatially varying reference field is the velocity and since
it is linear in the spatial coordinate, the zeroth-order kinetic equation,
eq.(\ref{zero-KE})\ becomes%
\begin{equation}
\partial _{t}^{(0)}f^{(0)}+\mathbf{v}\cdot \left( \mathbf{\nabla }%
^{(0)}u_{0i}\right) \frac{\partial }{\partial u_{0i}}f^{(0)}=-\nu \left(
\psi \right) (f^{\left( 0\right) }-\phi _{0})+\frac{1}{2}\zeta ^{\ast
}\left( \alpha \right) \nu \left( \psi \right) \frac{\partial }{\partial 
\mathbf{v}}\cdot \left( \mathbf{C}f^{(0)}\right) .  \label{f00}
\end{equation}%
or, writing this in terms of the peculiar velocity, 
\begin{equation}
\partial _{t}^{(0)}f^{(0)}+v_{y}\partial _{y}^{0}f^{(0)}-av_{y}\frac{%
\partial }{\partial C_{x}}f^{(0)}=-\nu \left( \psi \right) (f-\phi _{0})+%
\frac{1}{2}\zeta ^{\ast }\left( \alpha \right) \nu \left( \psi \right) \frac{%
\partial }{\partial \mathbf{v}}\cdot \left( \mathbf{C}f^{(0)}\right) .
\end{equation}%
Here, the second term on the left accounts for any explicit dependence of
the distribution on the coordinate $y$, aside from the implicit dependence
coming from $\mathbf{C}$. Since it is a zero-order derivative, it does not
act on the deviations $\delta \psi $. In terms of the peculiar velocity,
this becomes%
\begin{equation}
\partial _{t}^{(0)}f^{(0)}+\left( C_{y}+\delta u_{y}\right) \partial
_{y}^{0}f^{(0)}-aC_{y}\frac{\partial }{\partial C_{x}}f^{(0)}-a\delta u_{y}%
\frac{\partial }{\partial C_{x}}f^{(0)}=-\nu \left( \psi \right) (f-\phi
_{0})+\frac{1}{2}\zeta ^{\ast }\left( \alpha \right) \nu \left( \psi \right) 
\frac{\partial }{\partial \mathbf{C}}\cdot \left( \mathbf{C}f^{(0)}\right) .
\label{f001}
\end{equation}%
The first term on the left is evaluated using eq.(\ref{zero-t}) and the
zeroth order balance equations{%
\begin{eqnarray}
{\partial _{t}^{(0)}n} &=&0\;  \label{T0} \\
{\partial _{t}^{(0)}u}_{i}+a\delta u_{y}\delta _{ix} &=&0  \notag \\
{\partial _{t}^{(0)}T}+\frac{2}{Dnk_{B}}aP_{xy}^{(0)} &=&-\zeta ^{\ast
}\left( \alpha \right) \nu \left( \psi \right) T\;,  \notag
\end{eqnarray}%
and the assumption of normality}%
\begin{equation*}
\partial _{t}^{(0)}f^{(0)}=\left( \partial _{t}^{(0)}\delta n\right) \left( 
\frac{\partial }{\partial \delta n}f^{(0)}\right) +\left( \partial
_{t}^{(0)}\delta T\right) \left( \frac{\partial }{\partial \delta T}%
f^{(0)}\right) +\left( \partial _{t}^{(0)}\delta u_{i}\right) \left( \frac{%
\partial }{\partial \delta u_{i}}f^{(0)}\right) 
\end{equation*}%
{to give}%
\begin{eqnarray}
&&\left( -\zeta ^{\ast }\left( \alpha \right) \nu \left( \psi \right) T-%
\frac{2}{Dnk_{B}}aP_{xy}^{(0)}\right) \frac{\partial }{\partial T}%
f^{(0)}-aC_{y}\frac{\partial }{\partial C_{x}}f^{(0)}-a\delta u_{y}\left( 
\frac{\partial }{\partial C_{x}}f^{(0)}+\frac{\partial }{\partial \delta
u_{x}}f^{(0)}\right)   \label{f0} \\
&=&-\nu ^{\ast }\left( \alpha \right) \nu \left( \psi \right) (f^{(0)}-\phi
_{0})+\frac{1}{2}\zeta ^{\ast }\left( \alpha \right) \nu \left( \psi \right) 
\frac{\partial }{\partial \mathbf{C}}\cdot \left( \mathbf{C}f^{(0)}\right) ,
\notag
\end{eqnarray}%
where the temperature derivative is understood to be evaluated at constant
density. Here, the second term on the left in eq.(\ref{f001}) has been
dropped as neither eq.(\ref{f001}) nor the balance equations contain
explicit reference to the velocity field $u_{0}$, and so no explicit
dependence on the coordinate $y$ , thus justifying the assumption that such
dependence does not occur in $f^{\left( 0\right) }$. One can also assume
that $f^{\left( 0\right) }$ depends on $\delta u_{i}$ only through the
peculiar velocity, since in that case the term proportional to $\delta u_{y}$%
vanishes as well and there is no other explicit dependence on $\delta u_{y}$.

Equation (\ref{f0}) is closed once the pressure tensor is specified. Since
the primary goal here is to develop the transport equations for deviations
from the reference state, attention will be focused on the determination of
the pressure tensor and the heat flux vector. It is a feature of the simple
kinetic model used here that these can be calculated without determining the
explicit form of the distribution.

\subsubsection{The zeroth-order pressure tensor}

An equation for the pressure tensor can be obtained by multiplying this
equation through by $mC_{i}C_{j}$ and integrating over velocities. Using the
definition given in eq.(\ref{fluxBGK}), 
\begin{equation}
\left( -\zeta ^{\ast }\left( \alpha \right) \nu \left( \psi \right) T-\frac{2%
}{Dnk_{B}}aP_{xy}^{(0)}\right) \frac{\partial }{\partial T}%
P_{ij}^{(0)}+a\delta _{ix}P_{jy}^{(0)}+a\delta _{jx}P_{iy}^{(0)}=-\nu ^{\ast
}\left( \alpha \right) \nu \left( \psi \right) (P_{ij}^{(0)}-\delta
_{ij}nk_{B}T)-\zeta ^{\ast }\left( \alpha \right) \nu \left( \psi \right)
P_{ij}^{(0)},  \label{p0}
\end{equation}%
and of course there is the constraint that by definition $Tr\left( \mathsf{P}%
\right) =Dnk_{B}T$. It is interesting to observe that eqs.,(\ref{T0}) - (\ref%
{p0}) are identical with their steady-state counterparts when the
steady-state condition, $\zeta ^{(0)}T=\frac{2}{Dnk_{B}}aP_{xy}^{(0)}$, is
fulfilled. However, here the solution of these equations is needed for
arbitrary values of $\delta T$, $\delta n$ and $\delta \mathbf{u}$. Another
point of interest is that these equations are local in the deviations $%
\delta \psi $ so that they are exactly the same equations as describe
spatially homogeneous deviations from the reference state. As mentioned
above, this is the meaning of the zeroth-order solution to the
Chapman-Enskog expansion: it is the exact solution to the problem of uniform
deviations from the reference state. It is this exact solution which is
''localized'' to give the zeroth-order Chapman-Enskog approximation and not
the reference distribution, $f_{0}$, except in the rare cases, such as
equilibrium, when they coincide.

To complete the specification of the distribution, eqs. (\ref{f0}) and (\ref%
{p0}) must be supplemented by boundary conditions. The relevant
dimensionless quantity characterizing the strength of the nonequilibrium
state is the dimensionless shear rate defined as 
\begin{equation}
a^{\ast }\equiv a/\nu =a\frac{\left( D+2\right) \Gamma \left( D/2\right) }{%
8\pi ^{\left( D-1\right) /2}n\sigma ^{D}}\sqrt{\frac{m\sigma ^{2}}{k_{B}T}}.
\end{equation}%
It is clear that for a uniform system, the dimensionless shear rate becomes
smaller as the temperature rises so that we expect that in the limit of
infinite temperature, the system will behave as an inelastic gas without any
shear - i.e., in the homogeneous cooling state, giving the boundary condition%
\begin{equation}
\lim_{T\rightarrow \infty }\frac{1}{nk_{B}T}P_{ij}=\delta _{ij},
\end{equation}%
and in this limit, the distribution must go to the homogeneous cooling state
distribution. These boundary conditions can be implemented equivalently by
rewriting eqs.(\ref{s-low}) in terms of the inverse temperature, or more
physically the variable $a^{\ast }$, and the dimensionless pressure tensor $%
P_{ij}^{(\ast )}=\frac{1}{nk_{B}T}P_{ij}^{(0)}$ giving%
\begin{equation}
\left( \frac{1}{2}\zeta ^{\ast }\left( \alpha \right) +\frac{1}{D}a^{\ast
}P_{xy}^{(\ast )}\right) a^{\ast }\frac{\partial }{\partial a^{\ast }}%
P_{ij}^{(\ast )}=\frac{2}{D}a^{\ast }P_{xy}^{(\ast )}P_{ij}^{(\ast
)}-a^{\ast }\delta _{ix}P_{jy}^{(\ast )}-a^{\ast }\delta _{jx}P_{iy}^{(\ast
)}-\nu ^{\ast }\left( \alpha \right) (P_{ij}^{(\ast )}-\delta _{ij})
\label{P0-a}
\end{equation}%
and writing $f^{(0)}\left( \mathbf{C};\psi \right) =n\left( \frac{m}{2\pi
k_{B}T}\right) ^{D/2}g\left( \sqrt{\frac{m}{k_{B}T}}\mathbf{C};a^{\ast
}\right) $ 
\begin{eqnarray}
&&\left( \zeta ^{\ast }\left( \alpha \right) +\frac{2}{D}a^{\ast
}P_{xy}^{(\ast )}\right) a^{\ast }\frac{\partial }{\partial a^{\ast }}g+%
\frac{1}{D}a^{\ast }P_{xy}^{(\ast )}C_{i}\frac{\partial }{\partial C_{i}}%
g+a^{\ast }P_{xy}^{(\ast )}g-a^{\ast }C_{y}\frac{\partial }{\partial C_{x}}g
\label{P00} \\
&=&-\nu ^{\ast }\left( \alpha \right) \left( g-\exp \left(
-mC^{2}/k_{B}T\right) \right) ,  \notag
\end{eqnarray}%
with boundary condition $\lim_{a^{\ast }\rightarrow 0}P_{ij}^{\left( \ast
\right) }=\delta _{ij}$ and $\lim_{a^{\ast }\rightarrow 0}g=\exp \left(
-mC^{2}/k_{B}T\right) $. For practical calculations, it is more convenient
to introduce a fictitious time variable, $s$, and to express these equations
as 
\begin{eqnarray}
\frac{da^{\ast }}{ds} &=&\frac{1}{2}a^{\ast }\zeta ^{\ast }\left( \alpha
\right) +\frac{1}{D}a^{\ast 2}P_{xy}^{(\ast )}  \label{ss-hi} \\
\frac{\partial }{\partial s}P_{ij}^{(0)} &=&\frac{2}{D}a^{\ast
}P_{xy}^{(\ast )}P_{ij}^{(\ast )}-a^{\ast }\delta _{ix}P_{jy}^{(\ast
)}-a^{\ast }\delta _{jx}P_{iy}^{(\ast )}-\nu ^{\ast }\left( \alpha \right)
(P_{ij}^{(\ast )}-\delta _{ij})  \notag
\end{eqnarray}%
where the boundary condition is then $P_{ij}^{\left( \ast \right) }\left(
s=0\right) =\delta _{ij}$, and $a^{\ast }\left( s=0\right) =0$. The
distribution then satisfies%
\begin{equation}
\frac{\partial }{\partial s}g=-\frac{1}{D}a^{\ast }P_{xy}^{(\ast )}C_{i}%
\frac{\partial }{\partial C_{i}}g-a^{\ast }P_{xy}^{(\ast )}g+a^{\ast }C_{y}%
\frac{\partial }{\partial C_{x}}g-\nu ^{\ast }\left( \alpha \right) \left(
g-\exp \left( -mC^{2}/k_{B}T\right) \right)   \label{ss-hif}
\end{equation}%
with $\lim_{s\rightarrow 0}g=\exp \left( -mC^{2}/k_{B}T\right) $. These are
to be solved simultaneously to give $P_{ij}^{\left( \ast \right) }\left(
s\right) ,a^{\ast }\left( s\right) $ and $f^{\left( 0\right) }\left(
s\right) $ from which the desired curves $P_{ij}^{\left( \ast \right)
}\left( a^{\ast }\right) $ and $f^{\left( 0\right) }\left( a^{\ast }\right) $
are obtained.

Physically, if the gas starts at a very high temperature, it would be
expected to cool until it reached the steady state. It is easy to see that
the right hand sides of eqs.(\ref{ss-hi}) do in fact vanish in the steady
state so that the steady state represents a critical point of this system of
differential equations\cite{Nicolis}. In order to fully specify the curve $%
P_{ij}\left( T\right) $ and the distribution $f^{(0)}$ it is necessary to
integrate as well from a temperature below the steady state temperature.
Clearly, in the case of \emph{zero} temperature, one expects that the
pressure tensor goes to zero since this corresponds to the physical
situation in which the atoms stream at exactly the velocities predicted by
their positions and the macroscopic flow field. (Note that if the atoms have
finite size, this could still lead to collisions. However, the BGK kinetic
theory used here is properly understood as an approximation to the Boltzmann
theory appropriate for a low density gas in which the finite size of the
grains is of no importance.) Thus, the expectation is that the
zero-temperature limit will give%
\begin{equation}
\lim_{T\rightarrow 0}P_{ij}^{(0)}=0.
\end{equation}%
Then, in terms of a fictitious time parameters, one has 
\begin{eqnarray}
\frac{dT}{ds} &=&-\zeta ^{\ast }\left( \alpha \right) \nu \left( \psi
\right) T-\frac{2}{D}aTP_{xy}^{(\ast )}  \label{s-low} \\
\frac{\partial }{\partial s}P_{ij}^{(\ast )} &=&a\frac{2}{D}P_{xy}^{(\ast
)}P_{ij}^{(\ast )}-a\delta _{ix}P_{jy}^{(\ast )}-a\delta _{jx}P_{iy}^{(\ast
)}-\nu ^{\ast }\left( \alpha \right) \nu \left( \psi \right) (P_{ij}^{(\ast
)}-\delta _{ij})  \notag
\end{eqnarray}%
and for the distribution%
\begin{equation}
\frac{\partial }{\partial s}f^{(0)}=aC_{y}\frac{\partial }{\partial C_{x}}%
f^{(0)}-\nu ^{\ast }\left( \alpha \right) \nu \left( \psi \right)
(f^{(0)}-\phi _{0})+\frac{1}{2}\zeta ^{\ast }\left( \alpha \right) \nu
\left( \psi \right) \frac{\partial }{\partial \mathbf{C}}\cdot \left( 
\mathbf{C}f^{(0)}\right) .  \label{s-low1}
\end{equation}%
A final point is that the solution of these equations requires more than the
boundary condition $P_{ij}^{(0)}\left( s=0\right) =0$ since evaluation of
the right hand side of eq.(\ref{s-low}) requires a statement about $%
P_{ij}^{(\ast )}\left( s=0\right) $ as well. A straight-forward series
solution of eq.(\ref{p0}) in the vicinity of $T=0$  gives $P_{xy}^{\ast
}\sim a^{\ast -1/3}$ and $P_{ii}^{\ast }\sim a^{\ast -2/3}$so that the
correct boundary condition here is $P_{ij}^{(\ast )}\left( s=0\right) =0$.
The solution of these equations can then be performed as discussed in ref. %
\cite{SantosInherentRheology} with the boundary conditions given here.

It will also prove useful below to know the behavior of the pressure tensor
near the steady-state. This is obtained by making a series solution to eq.(%
\ref{P0-a}) in the variable $\left( a^{\ast }-a_{ss}^{\ast }\right) $ where $%
a_{ss}^{\ast }$ is the reduced shear in the steady-state. Details are given
in Appendix \ref{AppP} and the result is that%
\begin{equation}
P_{ij}^{\left( 0\right) }=P_{ij}^{ss}\left( 1+A_{ij}^{\ast }\left( \alpha
\right) \left( \frac{a^{\ast }}{a_{ss}^{\ast }}-1\right) +...\right) ,
\label{Pss}
\end{equation}%
with the coefficients%
\begin{eqnarray}
A_{xy}^{\ast }\left( \alpha \right)  &=&-2\frac{\Delta \left( \alpha \right)
+\zeta ^{\ast }\left( \alpha \right) }{\zeta ^{\ast }\left( \alpha \right) }
\label{Pss-A} \\
\left( 1-\delta _{ix}\right) A_{ii}^{\ast }\left( \alpha \right) 
&=&-2\left( \frac{\nu ^{\ast }\left( \alpha \right) +\zeta ^{\ast }\left(
\alpha \right) }{\Delta \left( \alpha \right) +\nu ^{\ast }\left( \alpha
\right) +\frac{1}{2}\zeta ^{\ast }\left( \alpha \right) }\right) \left(
1-\delta _{ix}\right)   \notag \\
A_{xx}^{\ast }\left( \alpha \right)  &=&-2D\frac{\left( \Delta \left( \alpha
\right) +\frac{1}{D}\nu ^{\ast }\left( \alpha \right) +\frac{1}{2}\zeta
^{\ast }\left( \alpha \right) \right) \left( \nu ^{\ast }\left( \alpha
\right) +\zeta ^{\ast }\left( \alpha \right) \right) }{\left( \Delta \left(
\alpha \right) +\nu ^{\ast }\left( \alpha \right) +\frac{1}{2}\zeta ^{\ast
}\left( \alpha \right) \right) \left( \nu ^{\ast }\left( \alpha \right)
+D\zeta ^{\ast }\left( \alpha \right) \right) },  \notag
\end{eqnarray}%
where $\Delta \left( \alpha \right) $ is the real root of 
\begin{equation}
4\Delta ^{3}+8\left( \nu ^{\ast }\left( \alpha \right) +\zeta ^{\ast }\left(
\alpha \right) \right) \Delta ^{2}+\left( 4\nu ^{\ast 2}\left( \alpha
\right) +14\nu ^{\ast }\left( \alpha \right) \zeta ^{\ast }\left( \alpha
\right) +7\zeta ^{\ast 2}\left( \alpha \right) \right) \Delta +\zeta ^{\ast
}\left( \alpha \right) \left( 2\nu ^{\ast 2}\left( \alpha \right) -\nu
^{\ast }\left( \alpha \right) \zeta ^{\ast }\left( \alpha \right) -2\zeta
^{\ast 2}\left( \alpha \right) \right) =0.  \label{Pss-d1}
\end{equation}

\subsubsection{Higher order moments:\ the zeroth-order heat flux vector}

Determination of the heat flux vector requires consideration of the full
tensor of third order moments. Since fourth order moments will also be
needed later, it is easiest to consider the equations for the general $N$-th
order moment, defined as 
\begin{equation}
M_{i_{1}...iN}^{(0)}\left( \mathbf{r,}t\right) =m\int d\mathbf{v}%
\;C_{i_{1}}...C_{i_{N}}f^{\left( 0\right) }\left( \mathbf{r,C,}t\right) .
\end{equation}%
To simplify the equations, a more compact notation will be used for the
indices whereby a collection of numbered indices, such as $i_{1}...i_{N}$,
will be written more compactly as $I_{N}$ so that capital letters denote
collections of indices and the subscript on the capital indicates the number
of indices in the collection. Some examples of this are%
\begin{eqnarray}
M_{I_{N}}^{(0)} &=&M_{i_{1}...i_{N}}^{(0)} \\
M_{I_{2}}^{(0)} &=&M_{i_{1}i_{2}}^{(0)}  \notag \\
M_{I_{2}y}^{(0)} &=&M_{i_{1}i_{2}y}^{(0)}.  \notag
\end{eqnarray}

In terms of the general moments, the heat flux vector is 
\begin{equation}
q_{i}^{\left( 0\right) }\left( \mathbf{r,}t\right) =\frac{1}{2}%
\sum_{j}M_{ijj}^{(0)}\left( \mathbf{r,}t\right) =\frac{1}{2}%
M_{ijj}^{(0)}\left( \mathbf{r,}t\right) ,
\end{equation}%
where the second equality introduces the Einstein summation convention
whereby repeated indices are summed. The pressure tensor is just the second
moment $P_{ij}^{\left( 0\right) }=M_{ij}^{\left( 0\right) }$. The local
equilibrium moments are easily shown to be zero for odd $N$ while the result
for even $N$ is%
\begin{equation}
M_{I_{N}}^{\left( le\right) }=mn\left( \frac{2k_{B}T}{m}\right) ^{\frac{N}{2}%
}2^{\frac{N}{2}}\frac{\Gamma \left( \frac{N+1}{2}\right) \Gamma \left( \frac{%
N+2}{2}\right) }{\sqrt{\pi }\Gamma \left( N\right) }\mathcal{P}%
_{I_{N}}\delta _{i_{1}i_{2}}\delta _{i_{3}i_{4}}...\delta _{i_{N-1}i_{N}}
\end{equation}%
where the operator $\mathcal{P}_{ijk...}$ indicates the sum over distinct
permutations of the indices $ijk...$ and has no effect on any other indices.
(e.g., $\mathcal{P}_{I_{4}}\delta _{i_{1}i_{2}}\delta _{i_{3}i_{4}}=\delta
_{i_{1}i_{2}}\delta _{i_{3}i_{4}}+\delta _{i_{1}i_{3}}\delta
_{i_{2}i_{4}}+\delta _{i_{1}i_{4}}\delta _{i_{2}i_{3}}$). An equation for
the general $N$-th order moment can be obtained from eq.(\ref{f00}) with the
result%
\begin{equation}
\left( -\zeta ^{\ast }\left( \alpha \right) -\frac{2}{D}a^{\ast
}P_{xy}^{(\ast )}\right) T\frac{\partial }{\partial T}M_{I_{N}}^{(0)}+\left(
\nu ^{\ast }\left( \alpha \right) +\frac{N}{2}\zeta ^{\ast }\left( \alpha
\right) \right) M_{I_{N}}^{(0)}+a^{\ast }\mathcal{P}_{I_{N}}\delta
_{xi_{N}}M_{I_{N-1}y}^{(0)}=\nu ^{\ast }\left( \alpha \right)
M_{I_{N}}^{(le)}.
\end{equation}%
Writing $M_{I_{N}}^{(0)}=mn\left( \frac{2k_{B}T}{m}\right) ^{\frac{N}{2}%
}M_{I_{N}}^{\ast }$gives%
\begin{equation}
-\left( \zeta ^{\ast }\left( \alpha \right) +\frac{2}{D}a^{\ast
}P_{xy}^{(\ast )}\right) T\frac{\partial }{\partial T}M_{I_{N}}^{\ast
}+\left( \nu ^{\ast }\left( \alpha \right) -\frac{N}{D}a^{\ast
}P_{xy}^{(\ast )}\right) M_{I_{N}}^{\ast }+a^{\ast }\mathcal{P}%
_{I_{N}}\delta _{xi_{N}}M_{I_{N-1}y}^{\ast }=\nu ^{\ast }\left( \alpha
\right) M_{I_{N}}^{(le\ast )}  \label{Moments1}
\end{equation}%
Notice that the moments are completely decoupled order by order in $N$.
Since the source on the right vanishes for odd $N$ it is natural to assume
that $M_{I_{N}}^{\ast }=0$ for odd $N$. This is certainly true for
temperatures above the steady state temperature since the appropriate
boundary condition in this case, based on the discussion above, is that $%
\lim_{T\rightarrow \infty }M_{I_{N}}^{\ast }=M_{I_{N}}^{(le\ast )}=0$. In
the opposite limit, $T\rightarrow 0$, as mentioned above, one has that $%
P_{xy}^{\ast }\sim a^{\ast -1/3}\sim T^{1/6}$ and there are two cases to
consider depending on whether or not the third term on the left contributes.
If it does, i.e. if one or more indices is equal to $x$, then a series
solution near $T=0$ gives $M_{I_{N}}^{\ast }\sim a^{\ast -1}\sim T^{1/2}$
while if no index is equal to $x$ then $M_{I_{N}}^{\ast }\sim a^{\ast
-2/3}\sim T^{1/3}$ giving in both cases the boundary condition $%
\lim_{T\rightarrow 0}M_{I_{N}}^{\ast }=0$. In particular, this shows that
the odd moments vanish for all temperatures. From this, it immediately
follows that 
\begin{equation}
q_{i}^{\left( 0\right) }\left( \mathbf{r,}t\right) =0.
\end{equation}

\subsection{First-order Chapman-Enskog: General formalism}

The equation for the first-order distribution, eq.(\ref{f1}), becomes%
\begin{equation}
\partial _{t}^{(0)}f^{(1)}+av_{y}\frac{\partial }{\partial u_{0x}}%
f^{(1)}=-\nu \left( \psi \right) f^{\left( 1\right) }+\frac{1}{2}\zeta
^{\ast }\left( \alpha \right) \nu \left( \psi \right) \frac{\partial }{%
\partial \mathbf{v}}\cdot \left( \mathbf{C}f^{(1)}\right) -\left( \partial
_{t}^{(1)}f^{(0)}+\mathbf{v}\cdot \mathbf{\nabla }_{1}f^{(0)}\right) ,
\end{equation}%
and the operator $\partial _{t}^{(1)}$ is defined via the corresponding
balance equations which are now {%
\begin{eqnarray}
{\partial _{t}^{(1)}\delta n+\mathbf{u}\cdot \mathbf{\nabla }}\delta n+n{%
\mathbf{\nabla }}\cdot \delta \mathbf{u} &=&0\; \\
{\partial _{t}^{(1)}\delta u_{i}+\mathbf{u}\cdot \mathbf{\nabla }\delta }%
u_{i}+(mn)^{-1}\partial _{j}^{\left( 1\right) }P_{ij}^{\left( 0\right)
}+(mn)^{-1}\partial _{y}^{\left( 0\right) }P_{iy}^{\left( 1\right) } &=&0 
\notag \\
{\partial _{t}^{(1)}\delta T+\mathbf{u}\cdot \mathbf{\nabla }\delta }T+\frac{%
2}{Dnk_{B}}\left( P_{ij}^{\left( 0\right) }\nabla _{j}\delta u_{i}+{\mathbf{%
\nabla }}^{\left( 0\right) }\cdot \mathbf{q}^{(1)}+aP_{xy}^{\left( 1\right)
}\right)  &=&0.  \notag
\end{eqnarray}%
}\newline
Writing the kinetic equation in the form%
\begin{eqnarray}
&&\partial _{t}^{(0)}f^{(1)}+a\frac{\partial }{\partial u_{0x}}%
v_{y}f^{(1)}+\nu ^{\ast }\left( \alpha \right) \nu \left( \psi \right)
f^{\left( 1\right) }-\frac{1}{2}\zeta ^{\ast }\left( \alpha \right) \nu
\left( \psi \right) \frac{\partial }{\partial \mathbf{v}}\cdot \left( 
\mathbf{C}f^{(1)}\right)  \\
&=&-\left( \partial _{t}^{(1)}n+u_{l}\partial _{l}^{1}n\right) \frac{%
\partial }{\partial n}f^{(0)}-\left( \partial _{t}^{(1)}T+u_{l}\partial
_{l}^{1}T\right) \frac{\partial }{\partial T}f^{(0)}-\left( \partial
_{t}^{(1)}\delta u_{j}+u_{l}\partial _{l}^{1}\delta u_{j}\right) \frac{%
\partial }{\partial \delta u_{j}}f^{(0)}  \notag \\
&&-\left( \partial _{l}^{1}u_{l}\right) f^{(0)}-\partial _{l}^{1}C_{l}f^{(0)}
\notag
\end{eqnarray}%
equations for the $N$-th moment can be obtained by multiplying through by $%
C_{i_{1}}...C_{i_{N}}$ and integrating over velocity. The first two terms on
the left contribute 
\begin{eqnarray}
\int C_{i_{1}}...C_{i_{N}}\left( \partial _{t}^{(0)}f^{(1)}+a\frac{\partial 
}{\partial u_{0x}}v_{y}f^{(1)}\right) d\mathbf{v} &=&\partial
_{t}^{(0)}M_{I_{N}}^{\left( 1\right) }\mathbf{+}\mathcal{P}_{I_{N}}\left(
\partial _{t}^{\left( 0\right) }\delta u_{i_{N}}\right) M_{I_{N-1}}^{\left(
1\right) } \\
&&+a\frac{\partial }{\partial u_{0x}}\left( M_{I_{N}y}^{\left( 1\right)
}+\delta u_{y}M_{I_{N}}^{\left( 1\right) }\right) \mathbf{+}a\mathcal{P}%
_{I_{N}}\delta _{xi_{N}}\left( M_{I_{N-1}y}^{\left( 1\right) }+\delta
u_{y}M_{I_{N-1}}^{\left( 1\right) }\right)   \notag \\
&=&\partial _{t}^{(0)}M_{I_{N}}^{\left( 1\right) }+a\frac{\partial }{%
\partial u_{0x}}\left( M_{I_{N}y}^{\left( 1\right) }+\delta
u_{y}M_{I_{N}}^{\left( 1\right) }\right) \mathbf{+}a\mathcal{P}%
_{I_{N}}\delta _{xi_{N}}M_{I_{N-1}y}^{\left( 1\right) }  \notag
\end{eqnarray}%
where the last line follows from using the zeroth order balance equation $%
\partial _{t}^{\left( 0\right) }\delta u_{i_{N}}=-a\delta _{i_{N}x}\delta
u_{y}$ . The evaluation of the right hand side is straightforward with the
only difficult term being%
\begin{equation}
\int C_{i_{1}}...C_{i_{N}}\left( \frac{\partial }{\partial \delta u_{j}}%
f^{(0)}\right) d\mathbf{v=}\frac{\partial }{\partial \delta u_{j}}%
M_{I_{N}}^{\left( 0\right) }\mathbf{+}\mathcal{P}_{I_{N}}\delta
_{i_{N}j}M_{I_{N-1}}^{\left( 0\right) },
\end{equation}%
and from eq.(\ref{Moments1}) it is clear that $M_{I_{N}}^{\left( 0\right) }$
is independent of $\delta u_{j}$ so that the first term on the right
vanishes. Thus%
\begin{eqnarray}
&&\partial _{t}^{(0)}M_{I_{N}}^{\left( 1\right) }+a\frac{\partial }{\partial
u_{0x}}\left( M_{I_{N}y}^{\left( 1\right) }+\delta u_{y}M_{I_{N}}^{\left(
1\right) }\right) \mathbf{+}a\mathcal{P}_{I_{N}}\delta
_{xi_{N}}M_{I_{N-1}y}^{\left( 1\right) }+\left( \nu ^{\ast }\left( \alpha
\right) +\frac{N}{2}\zeta ^{\ast }\left( \alpha \right) \right) \nu \left(
\psi \right) M_{I_{N}}^{\left( 1\right) }  \label{moments} \\
&=&-\left( \partial _{t}^{(1)}n+u_{l}\partial _{l}^{1}n\right) \frac{%
\partial }{\partial n}M_{I_{N}}^{\left( 0\right) }-\left( \partial
_{t}^{(1)}T+u_{l}\partial _{l}^{1}T\right) \frac{\partial }{\partial T}%
M_{I_{N}}^{\left( 0\right) }-\left( \partial _{t}^{(1)}\delta
u_{j}+u_{l}\partial _{l}^{1}\delta u_{j}\right) \mathcal{P}_{I_{N}}\delta
_{i_{N}j}M_{I_{N-1}}^{\left( 0\right) }  \notag \\
&&-\left( \partial _{l}^{1}u_{l}\right) M_{I_{N}}^{\left( 0\right) }-%
\mathcal{P}_{I_{N}}\left( \partial _{l}^{1}u_{i_{N}}\right)
M_{I_{N-1}l}^{\left( 0\right) }-\partial _{l}^{1}M_{I_{N}l}^{\left( 0\right)
}  \notag
\end{eqnarray}%
Superficially, it appears that the right hand side depends explicitly on the
reference field, since $u_{l}=u_{0.l}+\delta u_{l}$, which would in turn
generate an explicit dependence of the moments on the $y$-coordinate.
However, when the balance equations are used to eliminate $\partial
_{t}^{(1)}$ this becomes%
\begin{eqnarray}
&&\partial _{t}^{(0)}M_{I_{N}}^{\left( 1\right) }+a\frac{\partial }{\partial
u_{0x}}\left( M_{I_{N}y}^{\left( 1\right) }+\delta u_{y}M_{I_{N}}^{\left(
1\right) }\right) +a\mathcal{P}_{I_{N}}\delta _{xi_{N}}M_{I_{N-1}y}^{\left(
1\right) }+\left( \nu ^{\ast }\left( \alpha \right) +\frac{N}{2}\zeta ^{\ast
}\left( \alpha \right) \right) \nu \left( \psi \right) M_{I_{N}}^{\left(
1\right) } \\
&=&\left( \partial _{l}^{\left( 1\right) }\delta u_{l}\right) n\frac{%
\partial }{\partial n}M_{I_{N}}^{\left( 0\right) }+\frac{2}{Dnk_{B}}\left(
M_{lk}^{\left( 0\right) }\partial _{l}^{\left( 1\right) }\delta
u_{k}+aM_{xy}^{\left( 1\right) }\right) \frac{\partial }{\partial T}%
M_{I_{N}}^{\left( 0\right) }  \notag \\
&&+\frac{1}{mn}\mathcal{P}_{I_{N}}\left( \partial _{l}^{\left( 1\right)
}P_{li_{N}}^{\left( 0\right) }+\partial _{y}^{\left( 0\right)
}P_{yi_{N}}^{\left( 1\right) }\right) M_{I_{N-1}}^{\left( 0\right) }  \notag
\\
&&-\left( \partial _{l}^{1}\delta u_{l}\right) M_{I_{N}}^{\left( 0\right) }-%
\mathcal{P}_{I_{N}}\left( \partial _{l}^{1}\delta u_{i_{N}}\right)
M_{I_{N-1}l}^{\left( 0\right) }-\partial _{l}^{1}M_{I_{N}l}^{\left( 0\right)
}  \notag
\end{eqnarray}%
Then, assuming that the first-order moments are independent of the reference
field, $\mathbf{u}_{0}$, gives 
\begin{eqnarray}
&&\partial _{t}^{(0)}M_{I_{N}}^{\left( 1\right) }+a\mathcal{P}_{I_{N}}\delta
_{xi_{N}}M_{I_{N-1}y}^{\left( 1\right) }+\left( \nu ^{\ast }\left( \alpha
\right) +\frac{N}{2}\zeta ^{\ast }\left( \alpha \right) \right) \nu \left(
\psi \right) M_{I_{N}}^{\left( 1\right) }-\left( \frac{2a}{Dnk_{B}}\frac{%
\partial }{\partial T}M_{I_{N}}^{\left( 0\right) }\right) M_{xy}^{\left(
1\right) }  \label{moments2} \\
&=&\left[ \delta _{ab}\left( n\frac{\partial }{\partial n}M_{I_{N}}^{\left(
0\right) }-M_{I_{N}}^{\left( 0\right) }\right) +\frac{2}{Dnk_{B}}%
P_{ab}^{\left( 0\right) }\frac{\partial }{\partial T}M_{I_{N}}^{\left(
0\right) }-\mathcal{P}_{I_{N}}\delta _{bi_{N}}M_{I_{N-1}a}^{\left( 0\right) }%
\right] \left( \partial _{a}^{\left( 1\right) }\delta u_{b}\right)   \notag
\\
&&+\left[ \frac{1}{mn}\mathcal{P}_{I_{N}}\left( \frac{\partial }{\partial
\delta n}P_{li_{N}}^{\left( 0\right) }\right) M_{I_{N-1}}^{\left( 0\right) }-%
\frac{\partial }{\partial \delta n}M_{I_{N}l}^{\left( 0\right) }\right]
\left( \partial _{l}^{\left( 1\right) }\delta n\right)   \notag \\
&&+\left[ \frac{1}{mn}\mathcal{P}_{I_{N}}\left( \frac{\partial }{\partial
\delta T}P_{li_{N}}^{\left( 0\right) }\right) M_{I_{N-1}}^{\left( 0\right) }-%
\frac{\partial }{\partial \delta T}M_{I_{N}l}^{\left( 0\right) }\right]
\left( \partial _{l}^{\left( 1\right) }\delta T\right)   \notag
\end{eqnarray}%
which is consistent since no factors of $\mathbf{u}_{0}$ appear and since
the zeroth order moments are known to be independent of the reference
velocity field.

The moment equations are linear in gradients in the deviation fields, so
generalized transport coefficients via the definition%
\begin{equation}
M_{I_{N}}^{\left( 1\right) }=-\lambda _{I_{N}ab}\frac{\partial \delta \psi
_{b}}{\partial r_{a}}=-\mu _{I_{N}a}\frac{\partial \delta n}{\partial r_{a}}%
-\kappa _{I_{N}a}\frac{\partial \delta T}{\partial r_{a}}-\eta _{I_{N}ab}%
\frac{\partial \delta u_{a}}{\partial r_{b}}
\end{equation}%
where the transport coefficients for different values of $N$ have the same
name but can always be distinguished by the number of indices they carry.
The zeroth-order time derivative is evaluated using 
\begin{eqnarray}
\partial _{t}^{(0)}\lambda _{I_{N}ab}\frac{\partial \delta \psi _{b}}{%
\partial r_{a}} &=&\left( \partial _{t}^{(0)}\lambda _{I_{N}ab}\right) \frac{%
\partial \delta \psi _{b}}{\partial r_{a}}+\lambda _{I_{N}ab}\partial
_{t}^{(0)}\frac{\partial \delta \psi _{b}}{\partial r_{a}} \\
&=&\left( \left( \partial _{t}^{(0)}T\right) \frac{\partial \lambda
_{I_{N}ab}}{\partial T}+\left( \partial _{t}^{(0)}\delta u_{j}\right) \frac{%
\partial \lambda _{I_{N}ab}}{\partial \delta u_{j}}\right) \frac{\partial
\delta \psi _{b}}{\partial r_{a}}+\lambda _{I_{N}ab}\frac{\partial }{%
\partial r_{a}}\left( \partial _{t}^{(0)}\delta \psi _{b}\right)   \notag \\
&=&\left( \partial _{t}^{(0)}T\right) \frac{\partial \lambda _{I_{N}ab}}{%
\partial T}\frac{\partial \delta \psi _{b}}{\partial r_{a}}+\lambda
_{I_{N}ab}\frac{\partial \delta \psi _{c}}{\partial r_{a}}\frac{\partial
\left( \partial _{t}^{(0)}\delta \psi _{b}\right) }{\partial \delta \psi _{c}%
}
\end{eqnarray}%
where the third line follows from (a) the fact that the transport
coefficients will have no explicit dependence on the velocity field, as may
be verified from the structure of eq(\ref{moments2}) and (b) the fact that
the gradient here is a first order gradient $\nabla _{1}$ so that it only
contributes via gradients of the deviations of the fields thus giving the
last term on the right. Since the fields are independent variables, the
coefficients of the various terms $\frac{\partial \delta \psi _{b}}{\partial
r_{a}}$ must vanish independently. For the coefficients of the velocity
gradients, this gives%
\begin{eqnarray}
&&\left( \partial _{t}^{(0)}T\right) \frac{\partial }{\partial T}\eta
_{I_{N}ab}+\eta _{I_{N}ac}\frac{\partial \left( \partial _{t}^{(0)}\delta
u_{c}\right) }{\partial \delta u_{b}}+a\mathcal{P}_{I_{N}}\delta
_{xi_{N}}\eta _{I_{N-1}yab}+\left( \nu ^{\ast }\left( \alpha \right) +\frac{N%
}{2}\zeta ^{\ast }\left( \alpha \right) \right) \nu \left( \psi \right) \eta
_{I_{N}ab}-\left( \frac{2a}{Dnk_{B}}\frac{\partial }{\partial T}%
M_{I_{N}}^{\left( 0\right) }\right) \eta _{xyab} \\
&=&-\delta _{ab}\left( n\frac{\partial }{\partial n}M_{I_{N}}^{\left(
0\right) }-M_{I_{N}}^{\left( 0\right) }\right) -\frac{2}{Dnk_{B}}%
M_{ab}^{\left( 0\right) }\frac{\partial }{\partial T}M_{I_{N}}^{\left(
0\right) }+\mathcal{P}_{I_{N}}\delta _{bi_{N}}M_{I_{N-1}a}^{\left( 0\right)
}.  \notag
\end{eqnarray}%
The vanishing of the coefficients of the density gradients gives%
\begin{eqnarray}
&&\left( \partial _{t}^{(0)}T\right) \frac{\partial }{\partial T}\mu
_{I_{N}a}+\kappa _{I_{N}a}\frac{\partial \left( \partial _{t}^{(0)}T\right) 
}{\partial n}+a\mathcal{P}_{I_{N}}\delta _{xi_{N}}\mu
_{I_{N-1}ya}^{N}+\left( \nu ^{\ast }\left( \alpha \right) +\frac{N}{2}\zeta
^{\ast }\left( \alpha \right) \right) \nu \left( \psi \right) \mu
_{I_{N}a}-\left( \frac{2a}{Dnk_{B}}\frac{\partial }{\partial T}%
M_{I_{N}}^{\left( 0\right) }\right) \mu _{xya} \\
&&=-\frac{1}{mn}\mathcal{P}_{I_{N}}\left( \frac{\partial }{\partial \delta n}%
P_{ai_{N}}^{\left( 0\right) }\right) M_{I_{N-1}}^{\left( 0\right) }+\frac{%
\partial }{\partial \delta n}M_{I_{N}a}^{\left( 0\right) },  \notag
\end{eqnarray}%
while the vanishing of the coefficient of the temperature gradient gives

\begin{eqnarray}
&&\left( \partial _{t}^{(0)}T\right) \frac{\partial }{\partial T}\kappa
_{I_{N}a}+\frac{\partial \left( \partial _{t}^{(0)}T\right) }{\partial T}%
\kappa _{I_{N}a}+a\mathcal{P}_{I_{N}}\delta _{xi_{N}}\kappa
_{I_{N-1}ya}^{N}+\left( \nu ^{\ast }\left( \alpha \right) +\frac{N}{2}\zeta
^{\ast }\left( \alpha \right) \right) \nu \left( \psi \right) \kappa
_{I_{N}a}-\left( \frac{2a}{Dnk_{B}}\frac{\partial }{\partial T}%
M_{I_{N}}^{\left( 0\right) }\right) \kappa _{xya} \\
&&=-\frac{1}{mn}\mathcal{P}_{I_{N}}\left( \frac{\partial }{\partial \delta T}%
P_{ai_{N}}^{\left( 0\right) }\right) M_{I_{N-1}}^{\left( 0\right) }+\frac{%
\partial }{\partial \delta T}M_{I_{N}a}^{\left( 0\right) }.  \notag
\end{eqnarray}
Notice that for even moments, the source terms for the density and
temperature transport coefficients all vanish (as they involve odd
zeroth-order moments) and it is easy to verify that the boundary conditions
are consistent with $\mu _{I_{N}a}=\kappa _{I_{N}a}=0$ and only the velocity
gradients contribute. For odd values of $N$, the opposite is true and $\eta
_{I_{N}ab}=0$ while the others are in general nonzero.

\subsection{Navier-Stokes transport}

\subsubsection{The first order pressure tensor}

Specializing to the case $N=2$ gives the transport coefficients appearing in
the pressure tensor%
\begin{equation}
P_{I_{N}}^{\left( 1\right) }=-\eta _{I_{N}ab}\frac{\partial \delta u_{a}}{%
\partial r_{b}}
\end{equation}%
where the generalized viscosity satisfies%
\begin{eqnarray}
&&\left( \partial _{t}^{(0)}T\right) \frac{\partial }{\partial T}\eta
_{ijab}-a\eta _{ijax}\delta _{by}+a\delta _{xi}\eta _{jyab}+a\delta
_{xj}\eta _{iyab}+\left( \nu ^{\ast }\left( \alpha \right) +\zeta ^{\ast
}\left( \alpha \right) \right) \nu \left( \psi \right) \eta _{ijab}-\left( 
\frac{2a}{Dnk_{B}}\frac{\partial }{\partial T}P_{ij}^{\left( 0\right)
}\right) \eta _{xyab} \\
&=&-\delta _{ab}\left( n\frac{\partial }{\partial n}P_{ij}^{\left( 0\right)
}-P_{ij}^{\left( 0\right) }\right) -\frac{2}{Dnk_{B}}P_{ab}^{\left( 0\right)
}\frac{\partial }{\partial T}P_{ij}^{\left( 0\right) }+\delta
_{bi}P_{ja}^{\left( 0\right) }+\delta _{bj}P_{ia}^{\left( 0\right) }.  \notag
\end{eqnarray}

\subsubsection{First order third moments and the heat flux vector}

For the third moments, the contribution of density gradients to the heat
flux is well-known in the theory of granular fluids and the transport
coefficient is here the solution of 
\begin{eqnarray}
&&\left( \partial _{t}^{(0)}T\right) \frac{\partial }{\partial T}\mu _{ijka}+%
\frac{\partial \left( \partial _{t}^{(0)}T\right) }{\partial n}\kappa
_{ijka}+\left( \nu ^{\ast }\left( \alpha \right) +\frac{3}{2}\zeta ^{\ast
}\left( \alpha \right) \right) \nu \left( \psi \right) \mu _{ijka} \\
&&+a\delta _{xk}\mu _{ijya}+a\delta _{xi}\mu _{kjya}+a\delta _{xj}\mu _{ikya}
\notag \\
&=&-\frac{1}{mn}\left( \frac{\partial }{\partial n}P_{ak}^{\left( 0\right)
}\right) P_{ij}^{\left( 0\right) }-\frac{1}{mn}\left( \frac{\partial }{%
\partial n}P_{ai}^{\left( 0\right) }\right) P_{kj}^{\left( 0\right) }-\frac{1%
}{mn}\left( \frac{\partial }{\partial n}P_{aj}^{\left( 0\right) }\right)
P_{ik}^{\left( 0\right) }+\frac{\partial }{\partial n}M_{ijka}^{\left(
0\right) },  \notag
\end{eqnarray}%
and the generalized thermal conductivity is determined from

\begin{eqnarray}
&&\left( \partial _{t}^{(0)}T\right) \frac{\partial }{\partial T}\kappa
_{ijka}+\frac{\partial \left( \partial _{t}^{(0)}T\right) }{\partial T}%
\kappa _{ijka}+\left( \nu ^{\ast }\left( \alpha \right) +\frac{3}{2}\zeta
^{\ast }\left( \alpha \right) \right) \nu \left( \psi \right) \kappa _{ijka}
\\
&&+a\delta _{xk}\kappa _{ijya}+a\delta _{xi}\kappa _{kjya}+a\delta
_{xj}\kappa _{ikya}  \notag \\
&=&-\frac{1}{mn}\left( \frac{\partial }{\partial T}P_{ak}^{\left( 0\right)
}\right) P_{ij}^{\left( 0\right) }-\frac{1}{mn}\left( \frac{\partial }{%
\partial T}P_{ai}^{\left( 0\right) }\right) P_{kj}^{\left( 0\right) }-\frac{1%
}{mn}\left( \frac{\partial }{\partial T}P_{aj}^{\left( 0\right) }\right)
P_{ik}^{\left( 0\right) }+\frac{\partial }{\partial T}M_{ijka}^{\left(
0\right) }.  \notag
\end{eqnarray}%
Note that both of these require knowledge of the zeroth-order fourth
velocity moment $M_{ijka}^{\left( 0\right) }$. The heat flux vector is 
\begin{equation}
q_{i}^{\left( 1\right) }=-\overline{\mu }_{ia}\frac{\partial \delta n}{%
\partial r_{a}}-\overline{\kappa }_{ia}\frac{\partial \delta T}{\partial
r_{a}}
\end{equation}%
where%
\begin{eqnarray}
\overline{\mu }_{ia} &=&\mu _{ijja} \\
\overline{\kappa }_{ia} &=&\kappa _{ijja}.  \notag
\end{eqnarray}

\subsection{The second-order transport equations}

In this Section, the results obtained so far are put together so as to give
the Navier-Stokes equations for deviations from the steady state. The
Navier-Stokes equations result from the sum of the balance equations. To
first order, this takes the form{%
\begin{eqnarray}
{\partial _{t}n+\mathbf{u}\cdot \mathbf{\nabla }}\delta n+n{\mathbf{\nabla }}%
\cdot \delta \mathbf{u} &=&0\;  \label{first} \\
{\partial _{t}u_{i}+\mathbf{u}\cdot \mathbf{\nabla }\delta }u_{i}+a\delta
_{ix}\delta u_{y}+(mn)^{-1}\partial _{j}^{\left( 1\right) }P_{ij}^{\left(
0\right) } &=&0  \notag \\
{\partial _{t}T+\mathbf{u}\cdot \mathbf{\nabla }\delta }T+\frac{2}{Dnk_{B}}%
\left( P_{ij}^{\left( 0\right) }\nabla _{j}\delta
u_{i}+aP_{xy}^{(0)}+aP_{xy}^{\left( 1\right) }\right)  &=&-\zeta ^{\ast
}\left( \alpha \right) \nu \left( \psi \right) T.  \notag
\end{eqnarray}%
where }${\partial _{t}=\partial _{t}^{\left( 0\right) }+\partial
_{t}^{\left( 1\right) }}$. By analogy with the analysis of an equilibrium
system, these will be termed the Euler approximation. Summing to second
order to get the Navier-Stokes approximation gives{%
\begin{gather}
{\partial _{t}n+\mathbf{u}\cdot \mathbf{\nabla }}\delta n+n{\mathbf{\nabla }}%
\cdot \delta \mathbf{u}=0\;  \label{second} \\
{\partial _{t}u_{i}+\mathbf{u}\cdot \mathbf{\nabla }\delta }u_{i}+a\delta
_{ix}\delta u_{y}+(mn)^{-1}\partial _{j}^{\left( 1\right) }P_{ij}^{\left(
0\right) }+(mn)^{-1}\partial _{y}^{\left( 1\right) }P_{iy}^{\left( 1\right)
}+(mn)^{-1}\partial _{j}^{\left( 0\right) }P_{ij}^{\left( 2\right) }=0 
\notag \\
{\partial _{t}T+\mathbf{u}\cdot \nabla \delta }T+\frac{2}{Dnk_{B}}\left( {%
\mathbf{\nabla }}^{\left( 1\right) }\cdot \mathbf{q}^{(1)}+{\mathbf{\nabla }}%
^{\left( 0\right) }\cdot \mathbf{q}^{(2)}+P_{ij}^{\left( 0\right) }\nabla
_{j}\delta u_{i}+P_{ij}^{\left( 1\right) }\nabla _{j}\delta
u_{i}+aP_{xy}^{(0)}+aP_{xy}^{\left( 1\right) }+aP_{xy}^{\left( 2\right)
}\right) =-\zeta ^{\ast }\left( \alpha \right) \nu \left( \psi \right) T. 
\notag
\end{gather}%
where now }${\partial _{t}=\partial _{t}^{\left( 0\right) }+\partial
_{t}^{\left( 1\right) }+\partial _{t}^{\left( 2\right) }}$ but this
expression is problematic. Based on the results so far, it seems reasonable
to expect that $\partial _{j}^{\left( 0\right) }P_{ij}^{\left( 2\right) }={%
\mathbf{\nabla }}^{\left( 0\right) }\cdot \mathbf{q}^{(2)}=0$. However, to
consistently write the equations to third order requires knowledge of $%
P_{xy}^{\left( 2\right) }$ which is not available without extending the
solution of the kinetic equation to third order. The reason this problem
arises here, and not in the analysis about equilibrium, is that the shear
rate, $a$, arises from a gradient of the reference field. In the usual
analysis, such a term would be first order and $aP_{xy}^{\left( 2\right)
}=\left( \partial _{i}u_{0j}\right) P_{ij}^{\left( 2\right) }$would be of
third order and therefore neglected here. This is unfortunate and shows that
this method of analysis does not completely supplant the need to go beyond
the second-order solution in order to study shear flow. However, this
problem is not unique. In fact, in calculations of the transport
coefficients for the homogeneous cooling state of a granular gas, a similar
problem occurs in the calculation of the cooling rate: the true
Navier-Stokes expression requires going to third order in the solution of
the kinetic equation\cite{DuftyGranularTransport},\cite{LutskoCE}. (This is
because the source does not appear under a gradient, as can be seen in the
equations above.) Thus, it is suggested here that the same type of
approximation be accepted here, namely that the term $aP_{xy}^{\left(
2\right) }$ is neglected, so that the total pressure tensor and heat flux
vectors are%
\begin{eqnarray}
P_{ij} &=&P_{ij}^{\left( 0\right) }+P_{ij}^{\left( 1\right) } \\
q_{i} &=&q_{i}^{\left( 0\right) }+q_{i}^{\left( 1\right) }  \notag
\end{eqnarray}%
and the transport equations can be written as{%
\begin{eqnarray}
{\partial _{t}n+\mathbf{\nabla }}\cdot \left( n\mathbf{u}\right)  &=&0\;
\label{hydro-final} \\
{\partial _{t}u_{i}+\mathbf{u}\cdot \mathbf{\nabla }}u_{i}+(mn)^{-1}\partial
_{j}P_{ij} &=&0  \notag \\
{\partial _{t}T+\mathbf{u}\cdot \mathbf{\nabla }}T+\frac{2}{Dnk_{B}}\left( {%
\mathbf{\nabla }}\cdot \mathbf{q}+P_{ij}\nabla _{j}u_{i}\right)  &=&-\zeta
^{\ast }\left( \alpha \right) \nu \left( \psi \right) T.  \notag
\end{eqnarray}%
which is the expected form of the balance equations. The total fluxes are
given terms of the generalized transport coefficients}%
\begin{eqnarray}
P_{ij} &=&P_{ij}^{\left( 0\right) }-\eta _{ijab}\frac{\partial \delta u_{a}}{%
\partial r_{b}} \\
q_{i} &=&-\mu _{ijja}\frac{\partial \delta n}{\partial r_{a}}-\kappa _{ijja}%
\frac{\partial \delta T}{\partial r_{a}}.  \notag
\end{eqnarray}

\subsection{Linearized second-order transport}

Some simplification occurs if attention is restricted to the linearized form
of these equations. This is because, as noted in the previous Section,
several transport coefficients are proportional to $\delta u_{y}$ and
consequently do not contribute when the transport coefficients are
linearized. Taking this into account, the total fluxes are%
\begin{eqnarray}
P_{ij} &=&P_{ij}^{\left( ss\right) }+\left( \frac{\partial P_{ij}^{\left(
0\right) }}{\partial \delta n}\right) _{ss}\delta n+\left( \frac{\partial
P_{ij}^{\left( 0\right) }}{\partial \delta T}\right) _{ss}\delta T-\eta
_{ijab}^{ss}\frac{\partial \delta u_{a}}{\partial r_{b}} \\
q_{i} &=&-\overline{\mu }_{ia}^{ss}\frac{\partial \delta n}{\partial r_{a}}-%
\overline{\kappa }_{ia}^{ss}\frac{\partial \delta T}{\partial r_{a}},  \notag
\end{eqnarray}%
where the superscript on the transport coefficients, and subscript on the
derivatives, indicates that they are evaluated to zeroth order in the
deviations,%
\begin{equation*}
\left( \frac{\partial P_{ij}^{\left( 0\right) }}{\partial \delta n}\right)
_{ss}\equiv \lim_{\delta \psi \rightarrow 0}\frac{\partial P_{ij}^{\left(
0\right) }}{\partial \delta n},
\end{equation*}%
i.e. in the steady state. The defining expressions for the transport
coefficients simplify since the factor $\partial _{t}^{(0)}T$ is at least of
first order in the deviations from the steady state (since it vanishes in
the steady state) so that the temperature derivative can be neglected thus
transforming the differential equations into coupled algebraic equations.
Also, all remaining quantities are evaluated for $\delta \psi =0$, i.e. in
the steady state. Thus the viscosity becomes%
\begin{eqnarray}
&&-a_{ss}^{\ast }\eta _{ijax}^{ss}\delta _{by}+a_{ss}^{\ast }\delta
_{xi}\eta _{jyab}^{ss}+a_{ss}^{\ast }\delta _{xj}\eta _{iyab}^{ss}+\left(
\nu ^{\ast }\left( \alpha \right) +\zeta ^{\ast }\left( \alpha \right)
\right) \eta _{ijab}^{ss}-\frac{2a_{ss}^{\ast }}{Dn_{0}k_{B}}\left( \frac{%
\partial }{\partial T}P_{ij}^{\left( 0\right) }\right) _{ss}\eta _{xyab}^{ss}
\label{p-lin} \\
&=&-\nu ^{-1}\left( \psi _{0}\right) \delta _{ab}\left( n_{0}\left( \frac{%
\partial }{\partial n}P_{ij}^{\left( 0\right) }\right) _{ss}-P_{ij}^{\left(
ss\right) }\right) -\frac{2\nu ^{-1}\left( \psi _{0}\right) }{Dn_{0}k_{B}}%
P_{ab}^{\left( ss\right) }\left( \frac{\partial }{\partial T}P_{ij}^{\left(
0\right) }\right) _{ss}+\nu ^{-1}\left( \psi _{0}\right) \left( \delta
_{bi}P_{ja}^{\left( ss\right) }+\delta _{bj}P_{ia}^{\left( ss\right)
}\right)   \notag
\end{eqnarray}%
where $a_{ss}^{\ast }$ was defined in eq.(\ref{balance}). The generalized
heat conductivities will be given by the simplified equations 
\begin{eqnarray}
&&\nu ^{-1}\left( \psi _{0}\right) \left( \frac{\partial \left( \partial
_{t}^{(0)}T\right) }{\partial n}\right) _{ss}\kappa _{ijka}^{ss}+\left( \nu
^{\ast }\left( \alpha \right) +\frac{3}{2}\zeta ^{\ast }\left( \alpha
\right) \right) \mu _{ijka}^{ss}+a_{ss}^{\ast }\mathcal{P}_{ijk}\delta
_{xk}\mu _{ijya}^{ss} \\
&=&-\frac{\nu ^{-1}\left( \psi _{0}\right) }{mn_{0}}\mathcal{P}_{ijk}\left( 
\frac{\partial }{\partial n}P_{ak}^{\left( 0\right) }\right)
_{ss}P_{ij}^{\left( ss\right) }+\nu ^{-1}\left( \psi _{0}\right) \left( 
\frac{\partial }{\partial n}M_{ijka}^{\left( 0\right) }\right) _{ss}  \notag
\end{eqnarray}%
and

\begin{eqnarray}
&&\nu ^{-1}\left( \psi _{0}\right) \left( \frac{\partial \left( \partial
_{t}^{(0)}T\right) }{\partial T}\right) _{ss}\kappa _{ijka}^{ss}+\left( \nu
^{\ast }\left( \alpha \right) +\frac{3}{2}\zeta ^{\ast }\left( \alpha
\right) \right) \kappa _{ijka}^{ss}+a_{ss}^{\ast }\mathcal{P}_{ijk}\delta
_{xk}\kappa _{ijya}^{ss} \\
&=&-\frac{\nu ^{-1}\left( \psi _{0}\right) }{mn_{0}}\mathcal{P}_{ijk}\left( 
\frac{\partial }{\partial T}P_{ak}^{\left( 0\right) }\right)
_{ss}P_{ij}^{\left( ss\right) }+\nu ^{-1}\left( \psi _{0}\right) \left( 
\frac{\partial }{\partial T}M_{ijka}^{\left( 0\right) }\right) _{ss}.  \notag
\end{eqnarray}%
In these equations, the hydrodynamic variables $\psi _{0}$ must satisfy the
steady state balance condition, eq.(\ref{balance}). The various quantities
in these equations are known from the analysis of the zeroth order moments.
For example, from eq.(\ref{Pss}), one has that%
\begin{eqnarray}
\left( \frac{\partial P_{ij}^{\left( 0\right) }}{\partial T}\right) _{ss}
&=&-\frac{1}{2}T_{0}^{-1}P_{ij}^{ss}A_{ij}^{\ast }\left( \alpha \right)  \\
\left( \frac{\partial P_{ij}^{\left( 0\right) }}{\partial n}\right) _{ss}
&=&n_{0}^{-1}P_{ij}^{ss}\left( 1-A_{ij}^{\ast }\left( \alpha \right) \right) 
\notag \\
\nu ^{-1}\left( \psi _{0}\right) \left( \frac{\partial \left( \partial
_{t}^{(0)}T\right) }{\partial T}\right) _{ss} &=&-\frac{1}{2}\zeta ^{\ast
}\left( \alpha \right) \left( 1+A_{xy}^{\ast }\left( \alpha \right) \right) 
\notag
\end{eqnarray}%
where $A_{ij}^{\ast }\left( \alpha \right) $ was given in eq.(\ref{Pss-A})
and here, there is no summation over repeated indices. The derivatives of
higher order moments in the steady state can easily be given using the
results in Appendix \ref{AppP}.The linearized transport equations are{%
\begin{eqnarray}
{\partial _{t}\delta n+ay}\frac{\partial }{\partial x}\delta n+n_{0}\mathbf{%
\nabla }\cdot \delta \mathbf{u} &=&0\; \\
{\partial _{t}\delta u_{i}+{ay\frac{\partial }{\partial x}}\delta }%
u_{i}+a\delta u_{y}\delta _{ix}+(mn_{0})^{-1}\left( \left( \frac{\partial
P_{ij}^{\left( 0\right) }}{\partial n}\right) _{ss}\frac{\partial \delta n}{%
\partial r_{j}}+\left( \frac{\partial P_{ij}^{\left( 0\right) }}{\partial T}%
\right) _{ss}\frac{\partial \delta T}{\partial r_{j}}+\eta _{ijab}^{ss}\frac{%
\partial ^{2}\delta u_{a}}{\partial r_{j}\partial r_{b}}\right)  &=&0  \notag
\end{eqnarray}%
}%
\begin{eqnarray*}
&&{\partial _{t}\delta T+ay}\frac{\partial }{\partial x}{\delta T}+\frac{2}{%
Dn_{0}k_{B}}\left( \overline{\mu }_{ia}^{ss}\frac{\partial ^{2}\delta n}{%
\partial r_{i}\partial r_{a}}+\overline{\kappa }_{ia}^{ss}\frac{\partial
^{2}\delta T}{\partial r_{i}\partial r_{a}}+P_{ij}^{\left( ss\right) }\frac{%
\partial \delta u_{i}}{\partial r_{j}}+a\eta _{xyab}^{ss}\frac{\partial
\delta u_{a}}{\partial r_{b}}\right)  \\
&&+\frac{2a}{Dn_{0}^{2}k_{B}}\left( n_{0}\left( \frac{\partial
P_{xy}^{\left( 0\right) }}{\partial \delta n}\right) _{ss}-P_{xy}^{\left(
ss\right) }\right) \delta n+\frac{2a}{Dn_{0}k_{B}}\left( \frac{\partial
P_{xy}^{\left( 0\right) }}{\partial \delta T}\right) _{ss}\delta T \\
&=&-\frac{3}{2}\zeta ^{\ast }\left( \alpha \right) \nu \left( \psi
_{0}\right) \delta T-\zeta ^{\ast }\left( \alpha \right) \nu \left( \psi
_{0}\right) T_{0}\frac{\delta n}{n_{0}}.
\end{eqnarray*}%
where the fact that $\nu \left( \psi \right) \sim nT^{1/2}$ has been used.
These equations have recently been used by Garz\'{o} to study the stability
of the granular fluid under uniform shear flow\cite{garzo-2005-}.

\section{Conclusions}

In this paper, the extension of the Chapman-Enskog method to arbitrary
reference states has been presented. One of the key ideas is the separation
of the gradient operator into ''zeroth'' and ''first'' order operators that
help to organize the expansion. It is also important that the zeroth order
distribution be recognized as corresponding to the exact distribution for 
\emph{arbitrary} \emph{uniform} deviations of \emph{all} hydrodynamic fields
from the reference state. This distribution does not in general have
anything to do with the distribution in the reference state, except in the
very special case that the reference state itself is spatially uniform.

The method was illustrated by application to the paradigmatic non-uniform
system of a fluid undergoing uniform shear flow. In particular, the fluid
was chosen to be a granular fluid which therefore admits of a steady state.
The analysis was based on a particularly simple kinetic theory in order to
allow for illustration of the general concepts without the technical
complications involved in, e.g., using the Boltzmann equation. Nevertheless,
it should be emphasized that the difference between the present calculation
and that using the Boltzmann equation would be no greater than in the case
of an equilibrium fluid. The main difference is that with the simplified
kinetic theory, it is possible to obtain closed equations for the velocity
moments without having to explicitly solve for the distribution. When
solving the Boltzmann equation, the moment equations are not closed and it
is necessary to resort to expansions in orthogonal polynomials. In that
case, the calculation is usually organized somewhat differently: attention
is focussed on solving directly for the distribution but this is only a
technical point.(In fact, Chapman originally developed his version of the
Chapman-Enskog method using Maxwell's moment equations while Enskog based
his on the Boltzmann equation\cite{ChapmanCowling}. The methods are of
course equivalent.)

It is interesting to compare the hydrodynamic equations derived here to the
''standard'' equations for fluctuations about a uniform granular fluid. As
might be expected, the hydrodynamic equations describing fluctuations about
the state of uniform shear flow are more complex in some ways than are the
usual Navier-Stokes equations for a granular fluid, but the similarities
with the simpler case are perhaps more surprising. The complexity arises
from the fact that the transport coefficients do not have the simple spatial
symmetries present in the homogeneous fluid where, e.g., there is a single
thermal conductivity rather than the vector quantity that occurs here.
However, just as in homogeneous system, the heat flux vector still only
couples to density and temperature gradients and the pressure tensor to
velocity gradients so that the hydrodynamics equations, eq.(\ref{hydro-final}%
), have the same structure as the Navier-Stokes equations for the
homogeneous system.

An additional complication in the general analysis presented here is that
the zeroth-order pressure tensor and the transport coefficients are obtained
as the solution to partial differential equations in the temperature rather
than as simple algebraic functions. This requires that appropriate boundary
conditions be supplied which will, in general, depend on the particular
problem being solved. Here, in the high-temperature limit, the
non-equilibrium effects are of no importance and the appropriate boundary
condition on all quantities is that they approach their equilibrium values.
Boundary conditions must also be given at low temperature as the two domains
are separated by the steady-state which represents a critical point. At low
temperatures, there are no collisions and no deviations from the macroscopic
state so that all velocity moments go to zero thus giving the necessary
boundary conditions. A particularly simple case occurs when the hydrodynamic
equations are linearized about the reference state as would be appropriate
for a linear stability analysis. Then, the transport properties are obtained
as the solution to simple algebraic equations.

A particular simplifying feature of uniform shear flow is that the flow
field has a constant first gradient and, as a result, the moments do not
explicitly depend on the flow field. This will not be true for more complex,
nonlinear flow fields. However, the application of the methods discussed in
Section II should make possible an analysis similar to that given here for
the simple case of uniform shear flow.

\bigskip

\begin{acknowledgements}
I am grateful to Vicente Garz\'{o} and Jim Dufty for several useful discussions. This work was supportd in part by the European Space Agency
under contract number C90105.
\end{acknowledgements}

\appendix

\section{Steady-state temperature limit}

\label{AppP}

Recall that in the steady state

\begin{eqnarray}
P_{ii}^{\ast \left( ss\right) } &=&\frac{\nu ^{\ast }\left( \alpha \right)
+\delta _{ix}D\zeta ^{\ast }\left( \alpha \right) }{\nu ^{\ast }\left(
\alpha \right) +\zeta ^{\ast }\left( \alpha \right) } \\
P_{xy}^{\ast \left( ss\right) } &=&-\frac{a_{ss}^{\ast }\nu ^{\ast }\left(
\alpha \right) }{\left( \nu ^{\ast }\left( \alpha \right) +\zeta ^{\ast
}\left( \alpha \right) \right) ^{2}},  \notag
\end{eqnarray}%
and the explicit form of the steady-state condition, eq.(\ref{ss}) giving
the value of the reduced shear in the steady state, $a_{ss}^{\ast }$, is 
\begin{equation}
\frac{a_{ss}^{\ast 2}\nu ^{\ast }\left( \alpha \right) }{\left( \nu ^{\ast
}\left( \alpha \right) +\zeta ^{\ast }\left( \alpha \right) \right) ^{2}}=%
\frac{D}{2}\zeta ^{\ast }\left( \alpha \right) .
\end{equation}%
Assume that the stresses are analytic in $a^{\ast }$ so that near the
singularity 
\begin{equation*}
P_{ij}^{\left( \ast \right) }=P_{ii}^{\ast \left( ss\right) }+A_{ij}\left(
a^{\ast }-a_{ss}^{\ast }\right) +...
\end{equation*}%
They satisfy eq.(\ref{P0-a}),%
\begin{equation}
\left( \frac{1}{2}\zeta ^{\ast }\left( \alpha \right) +\frac{1}{D}a^{\ast
}P_{xy}^{(\ast )}\right) a^{\ast }\frac{\partial }{\partial a^{\ast }}%
P_{ij}^{(\ast )}=\frac{2}{D}a^{\ast }P_{xy}^{(\ast )}P_{ij}^{(\ast
)}-a^{\ast }\delta _{ix}P_{jy}^{(\ast )}-a^{\ast }\delta _{jx}P_{iy}^{(\ast
)}-\nu ^{\ast }\left( \alpha \right) (P_{ij}^{(\ast )}-\delta _{ij})
\end{equation}%
so%
\begin{eqnarray}
\left( \frac{1}{D}P_{xy}^{(\ast ss)}+\frac{1}{D}a_{ss}^{\ast }A_{xy}\right)
a_{ss}^{\ast }A_{ij} &=&\frac{2}{D}P_{xy}^{(\ast ss)}P_{ij}^{(\ast ss)}+%
\frac{2}{D}a_{ss}^{\ast }A_{xy}P_{ij}^{\left( \ast ss\right) }+\frac{2}{D}%
a_{ss}^{\ast }P_{xy}^{(\ast ss)}A_{ij}^{(\ast )} \\
&&-\delta _{ix}P_{jy}^{(\ast )}-\delta _{jx}P_{iy}^{(\ast )}-a_{ss}^{\ast
}\delta _{ix}A_{jy}-a_{ss}^{\ast }\delta _{jx}A_{iy}-\nu ^{\ast }\left(
\alpha \right) A_{ij}  \notag
\end{eqnarray}%
or%
\begin{equation}
\frac{1}{D}a_{ss}^{\ast 3}A_{xy}A_{ij}+a_{ss}^{\ast }\left( \nu ^{\ast
}\left( \alpha \right) +\frac{1}{2}\zeta ^{\ast }\left( \alpha \right)
\right) A_{ij}-\frac{2}{D}a_{ss}^{\ast 2}A_{xy}P_{ij}^{\ast
(ss)}+a_{ss}^{\ast 2}\delta _{ix}A_{jy}+a_{ss}^{\ast 2}\delta
_{jx}A_{iy}=\nu ^{\ast }\left( \alpha \right) \left( P_{ij}^{ss(\ast
)}-\delta _{ij}\right) 
\end{equation}%
In component form this is%
\begin{eqnarray}
\frac{1}{D}a_{ss}^{\ast 3}A_{xy}A_{yy}+a_{ss}^{\ast }\left( \nu ^{\ast
}\left( \alpha \right) +\frac{1}{2}\zeta ^{\ast }\left( \alpha \right)
\right) A_{yy}-\frac{2}{D}a_{ss}^{\ast 2}A_{xy}P_{yy}^{\ast (ss)} &=&\nu
^{\ast }\left( \alpha \right) \left( P_{yy}^{ss(\ast )}-1\right)  \\
\frac{1}{D}a_{ss}^{\ast 3}A_{xy}^{2}+a_{ss}^{\ast }\left( \nu ^{\ast }\left(
\alpha \right) +\frac{1}{2}\zeta ^{\ast }\left( \alpha \right) \right)
A_{xy}-\frac{2}{D}a_{ss}^{\ast 2}A_{xy}P_{xy}^{\ast (ss)}+a_{ss}^{\ast
2}A_{yy} &=&\nu ^{\ast }\left( \alpha \right) P_{xy}^{ss(\ast )}  \notag
\end{eqnarray}%
Substituting%
\begin{eqnarray}
A_{xy} &=&DC/a_{ss}^{\ast 2} \\
\left( 1-\delta _{ix}\right) A_{ii} &=&\left( 1-\delta _{ix}\right)
DB/a_{ss}^{\ast }  \notag \\
\sum_{ii}A_{ii} &=&0  \notag
\end{eqnarray}%
gives%
\begin{eqnarray}
BC+\left( \nu ^{\ast }\left( \alpha \right) +\frac{1}{2}\zeta ^{\ast }\left(
\alpha \right) \right) B-\frac{2}{D}CP_{yy}^{\ast (ss)} &=&\frac{1}{D}\nu
^{\ast }\left( \alpha \right) \left( P_{yy}^{\ast (ss)}-1\right)  \\
C^{2}+\left( \nu ^{\ast }\left( \alpha \right) +\frac{1}{2}\zeta ^{\ast
}\left( \alpha \right) \right) C-\frac{2}{D}a_{ss}^{\ast }CP_{xy}^{\ast
(ss)}+a_{ss}^{\ast 2}B &=&\frac{1}{D}\nu ^{\ast }\left( \alpha \right)
a_{ss}^{\ast }P_{xy}^{\ast (ss)}  \notag
\end{eqnarray}%
and the steady-state condition makes this%
\begin{eqnarray}
BC+\left( \nu ^{\ast }\left( \alpha \right) +\frac{1}{2}\zeta ^{\ast }\left(
\alpha \right) \right) B-\frac{2}{D}CP_{yy}^{\ast (ss)} &=&\frac{1}{D}\nu
^{\ast }\left( \alpha \right) \left( P_{yy}^{\ast (ss)}-1\right)  \\
C^{2}+\left( \nu ^{\ast }\left( \alpha \right) +\frac{3}{2}\zeta ^{\ast
}\left( \alpha \right) \right) C+a_{ss}^{\ast 2}B &=&-\frac{1}{2}\nu ^{\ast
}\left( \alpha \right) \zeta ^{\ast }\left( \alpha \right) .  \notag
\end{eqnarray}%
The solution is 
\begin{equation}
B=P_{yy}^{\ast (ss)}\frac{1}{D}\frac{2C-\zeta ^{\ast }\left( \alpha \right) 
}{C+\nu ^{\ast }\left( \alpha \right) +\frac{1}{2}\zeta ^{\ast }\left(
\alpha \right) }  \notag
\end{equation}%
with $C$ being the real root of%
\begin{equation}
4C^{3}+8\left( \nu ^{\ast }\left( \alpha \right) +\zeta ^{\ast }\left(
\alpha \right) \right) C^{2}+\left( 4\nu ^{\ast 2}\left( \alpha \right)
+14\nu ^{\ast }\left( \alpha \right) \zeta ^{\ast }\left( \alpha \right)
+7\zeta ^{\ast 2}\left( \alpha \right) \right) C+\zeta ^{\ast }\left( \alpha
\right) \left( 2\nu ^{\ast 2}\left( \alpha \right) -\nu ^{\ast }\left(
\alpha \right) \zeta ^{\ast }\left( \alpha \right) -2\zeta ^{\ast 2}\left(
\alpha \right) \right) =0
\end{equation}

Finally, it is useful to note that the full pressure has the expansion%
\begin{eqnarray}
P_{ij} &=&nk_{B}TP_{ij}^{\ast } \\
&=&n\left( k_{B}T_{ss}+x\left( \frac{\partial k_{B}T}{\partial a^{\ast }}%
\right) _{ss}+...\right) \left( P_{ij}^{\ast (ss)}+xA_{ij}+...\right)  
\notag \\
&=&P_{ij}^{ss}+n\left( k_{B}T_{ss}A_{ij}+P_{ij}^{\ast (ss)}\left( \frac{%
\partial k_{B}T}{\partial a^{\ast }}\right) _{ss}+...\right)   \notag \\
&=&P_{ij}^{ss}+xnk_{B}T_{ss}\left( A_{ij}-2a_{ss}^{\ast -1}P_{ij}^{\ast
(ss)}\right) +...  \notag
\end{eqnarray}%
This gives%
\begin{eqnarray}
P_{xy} &=&P_{xy}^{ss}+nk_{B}T_{ss}\left( \frac{D}{a_{ss}^{\ast 2}}%
C-2a_{ss}^{\ast -1}P_{xy}^{\ast (ss)}\right) \left( a^{\ast }-a_{ss}^{\ast
}\right) +... \\
&=&P_{xy}^{ss}+nk_{B}T_{ss}\frac{D}{a_{ss}^{\ast 2}}\left( C+\zeta ^{\ast
}\left( \alpha \right) \right) \left( a^{\ast }-a_{ss}^{\ast }\right) +... 
\notag \\
&=&P_{xy}^{ss}\left[ 1-2\left( \frac{C+\zeta ^{\ast }\left( \alpha \right) }{%
\zeta ^{\ast }\left( \alpha \right) }\right) \left( \frac{a^{\ast }}{%
a_{ss}^{\ast }}-1\right) +...\right]   \notag
\end{eqnarray}%
and%
\begin{eqnarray}
P_{yy} &=&P_{yy}^{ss}+nk_{B}T_{ss}\left( A_{yy}-2a_{ss}^{\ast
-1}P_{yy}^{\ast (ss)}\right) \left( a^{\ast }-a_{ss}^{\ast }\right) +... \\
&=&P_{yy}^{ss}+nk_{B}T_{ss}\left( Da_{ss}^{\ast -1}P_{yy}^{\ast (ss)}\frac{1%
}{D}\frac{2C-\zeta ^{\ast }\left( \alpha \right) }{C+\nu ^{\ast }\left(
\alpha \right) +\frac{1}{2}\zeta ^{\ast }\left( \alpha \right) }%
-2a_{ss}^{\ast -1}P_{yy}^{\ast (ss)}\right) \left( a^{\ast }-a_{ss}^{\ast
}\right) +...  \notag \\
&=&P_{yy}^{ss}\left[ 1+\left( \frac{2C-\zeta ^{\ast }\left( \alpha \right) }{%
C+\nu ^{\ast }\left( \alpha \right) +\frac{1}{2}\zeta ^{\ast }\left( \alpha
\right) }-2\right) \left( \frac{a^{\ast }}{a_{ss}^{\ast }}-1\right) +...%
\right]  \\
&=&P_{yy}^{ss}\left[ 1-2\left( \frac{\nu ^{\ast }\left( \alpha \right)
+\zeta ^{\ast }\left( \alpha \right) }{C+\nu ^{\ast }\left( \alpha \right) +%
\frac{1}{2}\zeta ^{\ast }\left( \alpha \right) }\right) \left( \frac{a^{\ast
}}{a_{ss}^{\ast }}-1\right) +...\right]   \notag
\end{eqnarray}%
and%
\begin{eqnarray}
P_{xx} &=&Dnk_{B}T_{ss}+\left( a^{\ast }-a_{ss}^{\ast }\right) Dnk_{B}\left( 
\frac{\partial T}{\partial a^{\ast }}\right) _{ss}-\left( D-1\right)
P_{yy}+... \\
&=&Dnk_{B}T_{ss}-2\left( \frac{a^{\ast }}{a_{ss}^{\ast }}-1\right)
Dnk_{B}T_{ss}-\left( D-1\right) P_{yy}+...  \notag \\
&=&Dnk_{B}T_{ss}-\left( D-1\right) P_{yy}^{ss}+2\left[ -Dnk_{B}T_{ss}+\left(
D-1\right) P_{yy}^{ss}\left( \frac{\nu ^{\ast }\left( \alpha \right) +\zeta
^{\ast }\left( \alpha \right) }{C+\nu ^{\ast }\left( \alpha \right) +\frac{1%
}{2}\zeta ^{\ast }\left( \alpha \right) }\right) \right] \left( \frac{%
a^{\ast }}{a_{ss}^{\ast }}-1\right)  \\
&=&P_{xx}^{ss}+2P_{xx}^{ss}\left[ \frac{-D+\left( D-1\right) \frac{\nu }{\nu
+x}\left( \frac{\nu +x}{C+\nu +\frac{1}{2}x}\right) }{D-\left( D-1\right) 
\frac{\nu }{\nu +x}}\right] \left( \frac{a^{\ast }}{a_{ss}^{\ast }}-1\right) 
\\
&=&P_{xx}^{ss}\left[ 1-2D\frac{\left( C+\frac{1}{D}\nu ^{\ast }\left( \alpha
\right) +\frac{1}{2}\zeta ^{\ast }\left( \alpha \right) \right) \left( \nu
^{\ast }\left( \alpha \right) +\zeta ^{\ast }\left( \alpha \right) \right) }{%
\left( C+\nu ^{\ast }\left( \alpha \right) +\frac{1}{2}\zeta ^{\ast }\left(
\alpha \right) \right) \left( \nu ^{\ast }\left( \alpha \right) +D\zeta
^{\ast }\left( \alpha \right) \right) }\left( \frac{a^{\ast }}{a_{ss}^{\ast }%
}-1\right) +...\right] 
\end{eqnarray}

The general moment equations are%
\begin{align}
& \frac{1}{2}\left( \zeta ^{\ast }\left( \alpha \right) +\frac{2}{D}a^{\ast
}P_{xy}^{(\ast )}\right) a^{\ast }\frac{\partial }{\partial a^{\ast }}%
M_{I_{N}}^{\ast }+\left( \nu ^{\ast }\left( \alpha \right) -\frac{N}{D}%
a^{\ast }P_{xy}^{(\ast )}\right) M_{I_{N}}^{\ast } \\
& +a^{\ast }\mathcal{P}\delta _{i_{N}x}M_{I_{N-1}y}^{\ast }=\nu ^{\ast
}\left( \alpha \right) M_{I_{N}}^{(le\ast )}.  \notag
\end{align}%
In the linear approximation, writing 
\begin{equation}
M_{I_{N}}^{\ast }=M_{I_{N}}^{\ast ss}+xA_{I_{N}}^{\ast }+...
\end{equation}%
gives%
\begin{eqnarray}
&&\frac{1}{D}\left( P_{xy}^{(\ast ss)}+a_{ss}^{\ast }A_{xy}^{(\ast )}\right)
a_{ss}^{\ast }A_{I_{N}}^{\ast }+\left( \nu ^{\ast }\left( \alpha \right) +%
\frac{N}{2}\zeta ^{\ast }\left( \alpha \right) \right) A_{I_{N}}^{\ast
}+a_{ss}^{\ast }\mathcal{P}\delta _{i_{N}x}A_{I_{N-1}y}^{\ast }  \notag \\
&=&\left( \frac{N}{D}a_{ss}^{\ast }A_{xy}^{(\ast )}+\frac{N}{D}P_{xy}^{(\ast
ss)}\right) M_{I_{N}}^{\ast ss}-\mathcal{P}\delta
_{i_{N}x}M_{I_{N-1}y}^{\ast ss}
\end{eqnarray}%
or%
\begin{eqnarray}
&&\left( \frac{1}{D}a_{ss}^{\ast 2}A_{xy}^{(\ast )}+\nu ^{\ast }\left(
\alpha \right) +\frac{\left( N-1\right) }{2}\zeta ^{\ast }\left( \alpha
\right) \right) A_{I_{N}}^{\ast }+a_{ss}^{\ast }\mathcal{P}\delta
_{i_{N}x}A_{I_{N-1}y}^{\ast }  \notag \\
&=&\frac{N}{D}\left( a_{ss}^{\ast }A_{xy}^{(\ast )}+P_{xy}^{(\ast
ss)}\right) M_{I_{N}}^{\ast ss}-\mathcal{P}\delta
_{i_{N}x}M_{I_{N-1}y}^{\ast ss}
\end{eqnarray}%
which, for $N>2$ is a simple linear equation for $A_{I_{N}}^{\ast }$.

\bigskip 
\bibliographystyle{prsty}
\bibliography{physics}

\end{document}